\documentstyle{article}

\begin{document}

\def\levelset{{\cal X}}
\def\fxx{{\bf x}}
\def\fxy{{\bf y}}
\def\real{I\!\!R}

\begin{center}
{\LARGE Retarded Electromagnetic Interaction and Dynamic }\\
{\LARGE Foundation of Classical Statistical Mechanics and}\\
{\LARGE Elimination of Reversibility Paradox of Time Reversal}\\
\vskip 0.2in
{\large Mei Xiaochun}\\
\vskip 0.2in
\par
(Department of Physics, Fuzhou University, Fuzhou, 350025, China, E-mail: fzbgk@pub3.fz.fj.cn )
\end{center}
\begin{abstract}
\end{abstract}
\par
It is proved in the paper that the non-conservative dissipative force and asymmetry of time reversal can be naturally introduced into classical statistical mechanics after retarded electromagnetic interaction between charged micro-particles is considered. In this way, the rational dynamic foundation for classical statistical physics can established and the revised Liouville's equation is obtained. The micro-canonical ensemble, the canonical ensemble, the distribution of near-independent subsystem, the distribution law of the Maxwell-Boltzmann and the Maxwell distribution of velocities are achieved directly from the Liouville's equation without using the hypothesis of equal probability. The micro-canonical ensemble is considered unsuitable as the foundation of equivalent state theory again, for most of equivalent states of isolated systems are not the states with equal probabilities actually. The reversibility paradoxes in the processes of non-equivalent evolutions of macro-systems can be eliminated completely, and the united description of statistical mechanics of equivalent and non-equivalent states is reached. The revised BBKGY series equations and hydromechanics equations are reduced, the non-equivalent entropy of general systems is defined and the principle of entropy increment of the non-equivalent entropy is proved at last.
\\
\\
{\large 1. The fundamental problems existing in classical statistical physics }\\
\par
Though classical statistical mechanics has been highly developed, its foundation has not yet been built up well up to now $^{(1)}$. The first problem is about the rationality of the equal probability hypothesis or the micro-canonical ensemble hypothesis£¬which is used as the foundation of equilibrium theory now. The hypothesis had got much criticism since it was put forward. In order to provide the hypothesis a rational base, Boltzmann raised the ergodic theory, proving that as long as a system was ergodic, the hypothesis of equal probability would be tenable. However, the study shows that the evolutions of systems can't be ergodic generally $^{(2)}$. So the hypothesis of equal probability can only be regarded as a useful work principle without strict proof. As for the non-equilibrium statistical systems, we have no united and perfect theory at present. We do not know how a system to transform from a non-equilibrium state to an equilibrium one at present and how to define the non-equivalent entropy of general systems. Besides, there still exists a so-called reversibility paradox in the theory. Though lots of researches have been done, the really rational solution still remains to be explored. 
\par
The reversibility paradox has been a long-standing problem. There exist two forms of the reversibility paradox at present. The first is the so-called Poincanre's recurrence. It was put forward by Zermelo in 1896 based on a theorem provided by Poincanre in 1890 $^{(3)}$. According to this form, a conservative system in a limited space would return to the infinitely nearing neighbor reign of its initial state. The basic ideal of the proof is described as follows. For a conservative system, the Liouville's theorem is tenable, so the volume of phase space is unchanged in the evolution process. Because the volume of phase space is limited, the system should be recurrent after a long enough time's evolution. That is to say, the process is reversible. The second form was put forward by Loschmidt in 1876. Loschmidt thought that any micro-dynamic motion equation did not change under time reversal, or the motion of any single micro-particle was reversible, so after the velocities of all particles in a macro-system are reversed at the same time, the system would evolve along the completely opposite direction. Therefore, the process would be reversible. However, in the evolution processes of isolated macro-systems, what can be observed is that the processes are always irreversible. Therefore, there exists the so-called reversibility paradox. Though a large numbers of explanations have been given up to now, none of them are satisfying. 
\\
\\
{\large 2. The Lorents retarded force and the basic hypothesis of classical statistical physics }\\
\par
It is well-known that there exist two kinds of motion equations, one is for single particles, the other is for a statistical systems composed of a larger numbers of particles. The so-called micro-equation of motion that Loschmidt mentioned about in his time was actually the Newtonian equation $md^2\vec{r}/{d}t^2=\vec{F}$. In that time quantum mechanics has not been found. Whether the Newtonian equation can keep unchanged under time reversal depends on the form of force $\vec{F}$. Only when $\vec{F}$ is conservative, it dose. In general situations when $\vec{F}$ is relative to time $t$ and momentum $\vec{p}$, the Newtonian equation can't keep unchanged under time reversal in general. In the common statistical systems composed of charged micro-particles, the interaction forces between micro-particles are electromagnetic forces. In the systems composed of neutral atoms and molecules, atoms and molecules can be regarded as electromagnetic dipole moments and quadrupole moments for their deformations caused by interactions and interaction between them can considered as one between electromagnetic dipole moments and quadrupole moments. In the paper, we only discuss interaction between charged particles, but the principle is the same for neutral atoms and molecules. 
\par
If the retarded interaction is not considered, the Lorentz force between two particles with charges $q$ and $q'$ as well as velocities $\vec{v}$ and $\vec{v}'$ is
\begin{equation}
\vec{F}={{qq'\vec{r}}\over{r^3}}+{{qq'\vec{v}'\times(\vec{v}\times\vec{r})}\over{c^2r^2}}
\end{equation}
In the current statistical physics, we use the formula to describe interactions between micro-particles. Because the force is unchanged when $\vec{v}\rightarrow-\vec{v}$ and $\vec{v}'\rightarrow-\vec{v}'$ under time reversal, it is impossible for us to introduce the irreversibility of time reversal into the theory to solve the problems in non-equivalent statistical mechanics. However, it should be noted that according to special relativity, instantaneous interaction does not exist. Electromagnetic interaction propagates in the speed of light. So retarded interaction should be considered between micro-particles in statistical systems. It is proved below that after retarded interaction is considered, the Lorentz forces can not keep unchanged under time reversal again and a rational dynamic foundation can be established for classical statistical mechanics and the reversibility paradox can resolved well.
\par
Let $t',\vec{r}',\vec{v}'$ and $\vec{a}'$ represent retarded time, coordinate, velocity and acceleration, $t,\vec{r},\vec{v}$ and represent non-retarded time, coordinate, velocity and acceleration. A particle with charge $q_j$, velocity $\vec{v}'_j$ and acceleration $\vec{a}'_j$ at space point $\vec{r}'_j(t')$ and time $t'$ would cause retarded potentials as follows at space point $\vec{r}_i(t)$ and time $t$ 
\begin{equation}
\varphi_{ij}={{q_j}\over{(1-{{\vec{\nu}'_j\cdot\vec{n}'_{ij}}\over{c}})r'_{ij}}}~~~~~~~~~~~~\vec{A}_{ij}={{q_j\vec{\nu}'_j}\over{c(1-{{\vec{\nu}'_j\cdot\vec{n}'_{ij}}\over{c}})r'_{ij}}}
\end{equation}
In the formula $\vec{r}'_{ij}(t,t')=\vec{r}_i(t)-\vec{r}'_j(t')$, $r'_{ij}=\mid\vec{r}'_{ij}\mid$, $\vec{n}'_{ij}=\vec{r}'_{ij}/{r}'_{ij}$. Let $v'_{jn}=\vec{v}'_j\cdot\vec{n}'_{jn}$, $a'_{jn}=\vec{a}'_j\cdot\vec{n}'_{ij}$£¬ the intensities of electromagnetic fields caused by $j$ particle at space point $\vec{r}_i$ and time $t$ are
\begin{equation}
\vec{E}'_{ij}={{q_j(1-{{v'^2_j}\over{c^2}})(\vec{n}'_{ij}-{{\vec{v}'_j}\over{c}})}\over{(1-{{\vec{v}'_j\cdot\vec{n}'_{ij}}\over{c}})^3r'^2_{ij}}}+{{q_j\vec{n}'_{ij}\times[(\vec{n}'_{ij}-{{\vec{v}'_j}\over{c}})\times\vec{a}'_j]}\over{c^2(1-{{\vec{v}'_j\cdot\vec{n}'_{ij}}\over{c}})^3r'_{ij}}}~~~~~~~~\vec{B}'_{ij}=\vec{n}'_{ij}\times\vec{E}'_{ij}
\end{equation}
We call the forces as the retarded Lorentz forces in which retarded interaction has been considered. Suppose there are $N$ particles in the system. Suppose the $i$ particle with charge $q_i$ at space point $\vec{r}_i(t)$ at time $t$ moves in velocity $\vec{v}_i$, the retarded Lorentz force acted on the $i$ particle caused by the $j$ particle is
$$\vec{F}'_{Rij}={{q_i{q}_j(1-{{v'^2_j}\over{c^2}})}\over{(1-{{v'_{jn}}\over{c}})^3{r}'^2_{ij}}}[\vec{n}_{ij}(1-{{\vec{v}_i\cdot\vec{v}'_j}\over{c^2}})-{{\vec{v}'_j}\over{c}}(1-{{v_{in}}\over{c}})]$$
$$+{{q_i{q}_j}\over{c^2(1-{{v'_{jn}}\over{c}})^3r'_{ij}}}\{\vec{n}_{ij}[a'_{jn}(1-{{\vec{v}_i\cdot\vec{v}'_j}\over{c^2}})-{{\vec{v}_i\cdot\vec{a}'_j}\over{c}}(1-{{v'_{jn}}\over{c}})]$$
\begin{equation}
-{{\vec{v}'_j{a}'_{jn}}\over{c}}(1-{{v_{in}}\over{c}})-\vec{a}'_j(1-{{v_{in}}\over{c}})(1-{{v'_{jn}}\over{c}})\}
\end{equation}
In the formula, $v_{in}=\vec{n}_{ij}\cdot\vec{v}_i$, $v'_{jn}=\vec{n}_{ij}\cdot\vec{v}'_j$, $a'_{jn}=\vec{n}'_{ij}\cdot\vec{a}'_j$. In order to do rational approximate calculation, let's accumulate the magnitude order of acceleration. According to formula $a=v^2/{r}'$, $a/{c}^2$ has the magnitude order of $1/{r}'$. Suppose hydrogen atom can be regarded as a harmonic oscillator with amplitude $b$ and angle frequency $\omega$, so the acceleration of oscillator is $b\omega^2$. The magnitude order of energy of hydrogen atom is $E=hc{/}\lambda=mv^2=mb^2\omega^2$. Suppose the wavelength of photons emitted by hydrogen atom is $\lambda=4\times{10}^{-7}{M}$, it can be calculated that $a/{c}^2=1/{r}'\simeq{60}$. But the magnitude order of distance between atoms is $r\simeq{10}^{-10}{M}$, $1/{r}\simeq{10}^{10}$. According to Eq.(4), the first item of the retarded Lorentz force directs ratio to $1/{r}^2$. So for the neighbor interaction, we have $a/{c}^2{r}=1/{r}r'=6\times{10}^{-9}/{r}^2<<{1}/{r}^2$, the item containing acceleration can be omitted. For the distance interaction, $a/{c}^2\geq{1}/{r}^2$, the items containing acceleration can not be omitted On the other hand, according to the Maxwell distribution of velocity, the average speed of hydrogen atoms is $\bar{v}=\sqrt{8kT/\pi{m}}$. Under common temperature $T=300K$, we have $\bar{v}=6.3\times{10}^6{M}$, $\bar{v}/{c}=2.1\times{10}^{-2}$. So the item containing $v/{c}$ is much bigger than the item containing acceleration, even the items containing $v^2/{c}^2$ are bigger than that containing acceleration in near neighbor interaction. So we should retain the items containing $v/{c}$, $v^2/{c}^2$ and $va/{c}^3$, but omitted high order items, then write Eq.(4) as 
$$\vec{F}_{Rij}={{q_i{q}_j}\over{r'^2_{ij}}}[\vec{n}'_{ij}(1+{{3v'_{jn}}\over{c}}+{{6v'^2_{jn}}\over{c^2}}-{{v'^2_j}\over{c^2}}-{{\vec{v}_i\cdot\vec{v}'_j}\over{c^2}})-{{\vec{v}'_j}\over{c}}(1-{{v'_{in}}\over{c}}+{{3v'_{jn}}\over{c}})]$$
\begin{equation}
+{{q_i{q}_j}\over{c^2{r}'_{ij}}}\{\vec{n}_{ij}[a'_{ij}(1+{{3v'_{jn}}\over{c}})-{{\vec{v}_i\cdot\vec{a}'_j}\over{c}}]-{{\vec{v}'_j{a}'_{jn}}\over{c}}-\vec{a}'_j(1-{{v_{in}}\over{c}}+{{2v'_{jn}}\over{c}})\}
\end{equation}
For the convenience of discussion later, we write $\vec{F}'_{Rij}=\vec{F}'_{0ij}+\vec{F}'_{ij}$, $\vec{F}'_{0ij}=q_i{q}_j\vec{n}'_{ij}/r'^2_{ij}$ represent conservative part and $\vec{F}'_{ij}$ represents non-conservative part. The total force acted on the i-particle caused by the other particles is $\vec{F}'_{Ri}=\sum\vec{F}'_{0ij}+\sum\vec{F}'_{ij}=\vec{F}'_{0i}+\vec{F}'_i$. For the macro-systems composed of large numbers of neutral atoms and molecular, we can consider atoms and molecular as electromagnetic dipole moments and quadrupoles and obtain the retarded Lorents forces in the same way. 
\par
On the other hand, according to classical electromagnetic theory, there exists the radiation damping forces $\vec{G}$ acted on the accelerated particles. The radiation damping forces are relative to the acceleration of acceleration. Suppose $l$ is the dimension of the particles, particle's speed $v<<{c}$, acceleration $a<<{c}^2/{l}$, the acceleration of acceleration $\dot{a}<<{c}^3/{l}^2$, when the distribution of particle's charge is with spherical symmetry, we have $\vec{G}=\kappa\dot{\vec{a}}$, $\kappa=2q^2/{3}c^3$.
\par
In classical electromagnetic theory, the Hamiltonian equations of motions can be described by using the space coordinates and common momentums $\vec{p}_i$, or by using the spaces coordinates and canonical momentums $\vec{p}_{zi}$ with relation $\vec{p}_{zi}=\vec{p}_i+q_i\vec{A}_i/c$. Both are equivalent. In statistical mechanics, it is more convenient to use common momentums. The accelerations and the accelerations of accelerations of particles should be regarded as the function of coordinate, momentum and time. On the other hand, the acceleration of the -particle is caused by other particle's interactions, and it is relative to the retarded speeds and distances, i,e., $\vec{a}'_i=\vec{a}'_i(\vec{r}_i,\vec{p}_i,\vec{r}'_j,\vec{p}'_j)$, $\dot{\vec{a}}'_i=\dot{\vec{a}}'_i(\vec{r}_i,\vec{p}_i,\vec{r}'_j,\vec{p}'_j)$. So the total Hamiltonian of the non-conservative system composed of $N$ charged particles can be written as $H=H_0+H'$, in which $H_0$ is the conservative Hamiltonian
\begin{equation}
H_0=\sum_{i}{{\vec{p}^2_i}\over{2m_i}}+\sum_{i<j}{U}_{0ij}(r'_{ij})
\end{equation}
Here $U_{ij}(r'_{ij})$ is the conservative interaction energy, $H'$ is the non-conservative Hamiltonian
\begin{equation}
H'=\sum_{i}{U}_i(\vec{r}_i,t)+\sum_{i<j}{U}_{ij}(r'_{ij},\vec{p}_i,\vec{p}'_j)
\end{equation}
$U_i(\vec{r}_i,t)$ is the interaction caused by external force, $U_{ij}(r'_{ij},\vec{p}_i,\vec{p}'_j)$ is the interaction caused by non-conservative force. The motion equation of the i-particle is
\begin{equation}
\dot{x}_{i\sigma}={{\partial{H}_0}\over{\partial{p}_{i\sigma}}}~~~~~~~~~~~~~~~~~~~\dot{p}_{i\sigma}=-{{\partial{H}_0}\over{\partial{x}_{i\sigma}}}+Q'_{i\sigma}=F'_{0i\sigma}+Q'_{i\sigma}
\end{equation}
\begin{equation}
Q'_{i\sigma}=F_{ei\sigma}(\vec{r}_i,t)+F'_{i\sigma}(\vec{r}_i,\vec{r}'_j,\vec{p}_i,\vec{p}'_j)+G'_{i\sigma}(\vec{r}_i,\vec{r}'_j,\vec{p}_i,\vec{p}'_j)
\end{equation}
In the formula, $F'_{0i\sigma}$ is conservative force $Q'_{i\sigma}$ is total of non-conservation force, $F_{ei\sigma}$ is external force, $F'_{i\sigma}$ is total retarded non-conservation force, $G'_{i\sigma}$ is radiation damping forces. Because acceleration of acceleration of a particle is caused by other particles, we can write $G'_{i\sigma}=\sum_{j\neq{i}}{G}'_{ij\sigma}$ in general, $G'_{ij\sigma}$ contains the $j$ -particle's inference on the acceleration of $i$ -particle's particle. We can express $\vec{r}'$, $r'$, $\vec{v}'$ and $\vec{a}'$ by $\vec{r}$, $r$, $\vec{v}$ and $\vec{a}$ as shown in Eq.(35)-(38). Because when $t\rightarrow-t$, we have $\vec{v}\rightarrow-\vec{v}$, $\vec{a}\rightarrow\vec{a}$, so that the retarded Lorentz forces and the radiation damping forces can not keep unchanged under time reversal, and irreversibility would appear in theory. These will further be shown in detail later.
\par
Two basic hypotheses can be regarded as the foundation of classical statistical physics as shown below.
\par
1. The interaction forces between charged micro-particles in the macro-systems are the retarded Lorentz forces and the radiation damping forces showing in Eq.(9). For neutral atoms and molecular, the interactions forces can be considered as the retarded Lorentz forces between electromagnetic dipole moments and quadrupoles. 
\par
2. A classical statistical system can be described by the normalized distribution function of ensemble probability density $\rho=\rho(x_{i\sigma},p_{i\sigma},t)$. The average value of the physical quantity $u$ in the ensemble is
\begin{equation}
\bar{u}=\int{u}(x_{i\sigma},p_{i\sigma},t)\rho(x_{i\sigma},p_{i\sigma},t)\prod^{N}_{i=1}\prod^{3}_{\sigma=1}{d}x_{i\sigma}{d}p_{i\sigma}
\end{equation}
\par
We will establish the dynamic equations of classical statistical physics based on these two hypotheses, then discuss the problems of equivalent and non-equivalent states below.
\\
\\
{\large 3. The basic dynamic equation of classical statistical physics }\\
\par
After the retarded Lorentz force and radiation damping force are introduced, the time rate of change of the distribution function of ensemble probability density $\rho$ is
\begin{equation}
{{d\rho}\over{dt}}={{\partial\rho}\over{\partial{t}}}+\sum_{i\sigma}[{{\partial\rho}\over{\partial{x}_{i\sigma}}}\dot{x}_{i\sigma}+{{\partial\rho}\over{\partial{p}_{i\sigma}}}\dot{p}_{i\sigma}]={{\partial\rho}\over{\partial{t}}}+\sum_{i\sigma}[{{\partial\rho}\over{\partial{x}_{i\sigma}}}\dot{x}_{i\sigma}+{{\partial\rho}\over{\partial{p}_{i\sigma}}}(F_{ei\sigma}+F'_{0i\sigma}+F'_{i\sigma}+G'_{i\sigma})]
\end{equation}
By using Eq.(6) and the continuity equation
\begin{equation}
{{\partial\rho}\over{\partial{t}}}+\sum_{i\sigma}[{{\partial(\rho\dot{x}_{i\sigma})}\over{\partial{x}_{i\sigma}}}+{{\partial(\rho\dot{p}_{i\sigma})}\over{\partial{p}_{i\sigma}}}]=0
\end{equation}
Eq.(11) can be written as
\begin{equation}
{{d\rho}\over{dt}}=-\rho\sum_{i\sigma}({{\partial\dot{x}_{i\sigma}}\over{\partial{x}_{i\sigma}}}+{{\partial\dot{p}_{i\sigma}}\over{\partial{p}_{i\sigma}}})=-\rho\sum_{i\sigma}({{\partial{F}'_{i\sigma}}\over{\partial{p}_{i\sigma}}}+{{\partial{G}'_{i\sigma}}\over{\partial{p}_{i\sigma}}})
\end{equation}
From Eq.(4) we can obtain
\begin{equation}
{{\partial{F}'_{ij\sigma}}\over{\partial{p}_{i\sigma}}}=0
\end{equation}
Put Eq.(14) into Eq.(12), we get
\begin{equation}
{{d\rho}\over{dt}}={{\partial\rho}\over{\partial{t}}}+\sum_{i\sigma}[{{p_{i\sigma}}\over{m_i}}{{\partial\rho}\over{\partial{x}_{i\sigma}}}+(F_{ei\sigma}+F'_{0i\sigma}+F'_{i\sigma}+G'_{i\sigma}){{\partial\rho}\over{\partial{p}_{i\sigma}}}]=-\rho\sum_{i\sigma}{{\partial{G}'_{i\sigma}}\over{\partial{p}_{i\sigma}}}
\end{equation}
But $\partial{G}'_{i\sigma}/\partial{r}_{i\sigma}\neq{0}$ in general, for example, for a charged oscillating dipole, we have $x=A\sin(\omega{t}+\delta)$, $p=m\dot{x}=Am\omega\cos(\omega{t}+\delta)$, $\dot{a}=-A\omega^3\cos(\omega{t}+\delta)=-\omega^2{p}/{m}$, so we have $\partial{G}/\partial{p}=-\kappa\omega^2/{m}\neq{0}$ when the retarded effect is not considered. Eq.(15) can also be written as
\begin{equation}
{{\partial\rho}\over{\partial{t}}}+\sum_{i\sigma}[{{p_{i\sigma}}\over{m_i}}{{\partial\rho}\over{\partial{x}_{i\sigma}}}+(F_{ei\sigma}+F'_{0i\sigma}+F'_{i\sigma}){{\partial\rho}\over{\partial{p}_{i\sigma}}}]=-{{\partial(G'_{i\sigma}\rho)}\over{\partial{p}_{i\sigma}}}
\end{equation}
\par
    The Eq.(15) or (16) are just the motion equations the distribution function of ensemble probability density $\rho$ satisfy after the retarded Lorentz forces and radiation damping forces are introduced, it can be regarded as the basic equation of the classical statistical mechanics. When $F'_{i\sigma}=G'_{i\sigma}=0$, the equation becomes the current Liuve equation. Because $F'_{i\sigma}$ and $G'_{i\sigma}$ can not keep unchanged under time reversal, Eq.(15) and (16) also can't be unchanged under time reversal.
\\
\\
{\large 4. The statistical distribution of equivalent states }\\
\par
Let's first discuss the definition of equilibrium states and the hypothesis of equal probability in statistical physics. The equilibrium states in thermodynamics have strict definition. The equilibrium states are the states that the system's macro-natures do not change with time without external influence. But in classical statistical theory, the equilibrium states have no independent and strict definition. For the equilibrium states of isolated systems, the definition depends on the equal probability hypothesis. The hypothesis is that for the equilibrium states of isolated systems, the ensemble probability density is a constant between the curved surfaces of energies $E$ and $E+\Delta{E}$ in phase space. So in current theory, the equilibrium states of isolated systems are equivalent to the states of equal probability. For the equilibrium states of non-isolated systems, there is no strict definition now. Some documents consider them as the states that the average values of physical quantities do not change with time. But this kind of definition depends on what kinds of physical quantities are used. So it is improper, for the equilibrium is only the nature of system itself, it has nothing to do with other physical quantities. As for the hypothesis of equal probability, it is a priori hypothesis and can't be proved in theory. The reason for the hypothesis to exist may be based on so-called " the law of non-difference". The hypothesis had got much criticism since it was put forward. In order to provide the hypothesis a rational base, Boltzmann raised the ergodic theory, proving that as long as a system was ergodic, the hypothesis of equal probability would be tenable. However, the study shows that the evolutions of systems can't be ergodic generally $^{(2)}$. So the ergodic hypothesis can't be used as the foundation of classical statistical physics and we have to look for other way out.
\par
For this purpose, let's discuss the definition of equilibrium states in statistical physics at first. We define the equilibrium states of isolated systems as the states that there are no external forces acting on the systems and the ensemble probabilities do not change with time. This definition coincides with the definition of thermodynamics of equilibrium states. According to the definition, for isolated systems we have $\vec{F}_{ei}(\vec{r},t)=0$. When the equilibrium states are reached, we have $d\rho{/}{d}t=0$  and $\partial\rho{/}\partial{t}=0$. So in the light of Eq.(15) in equilibrium states, we have 
\begin{equation}
\sum_{i\sigma}{{\partial{G}'_{i\sigma}}\over{\partial{p}_{i\sigma}}}=0~~~~~~~~~~\sum_{i\sigma}[{{p_{i\sigma}}\over{m_i}}{{\partial\rho}\over{\partial{x}_{i\sigma}}}+(F_{ei\sigma}+F'_{0i\sigma}+F'_{i\sigma}+G'_{i\sigma}){{\partial\rho}\over{\partial{p}_{i\sigma}}}]=0
\end{equation}
They are just the equations that the ensemble probability density functions satisfy in the equilibrium states. There are many kinds of equilibrium states of isolated systems as shown below. 
\par
1. The particles are free in the systems with $\vec{F}'_{0i}=\vec{F}'_i=\vec{G}'_i=0$ and the ensemble probability density satisfies $\partial\rho{/}\partial{x}_{i\sigma}=0$ and $\partial\rho{/}\partial{p}_{i\sigma}=0$. In this case, the ensemble probability density function $\rho=$ constant, i.e., $\rho$ has nothing to do with $x_{i\sigma}$, $p_{i\sigma}$ and $t$. This is just the state of equal probability or the so-called micro-canonical ensemble in the current theory. 
\par
2. Particles in systems are only acted by conservative forces, $\vec{F}'_{0i}\neq{0}$, $\vec{F}'_i=\vec{G}'_i=0$, and $\partial\rho{/}\partial{x}_{i\sigma}\neq{0}$£¬ $\partial\rho{/}\partial{p}_{i\sigma}\neq{0}$. The retarded effect is not considered. In this case, we have
\begin{equation}
{{\partial\rho}\over{\partial{x}_{i\sigma}}}{{p_{i\sigma}}\over{m_i}}+{{\partial\rho}\over{\partial{p}_{i\sigma}}}{F}_{i\sigma}={{\partial\rho}\over{\partial{x}_{i\sigma}}}{{\partial{H}_0}\over{\partial{p}_{i\sigma}}}-{{\partial\rho}\over{\partial{p}_{i\sigma}}}{{\partial{H}_0}\over{\partial{x}_{i\sigma}}}=0
\end{equation}
In this kinds of equilibrium states, the ensemble probability density functions do not change with time, but the probabilities are different in the different points of phase space, or the equivalent states are with unequal probabilities. It is easy to be verified that the solution of Eq.(18) is
\begin{equation}
\rho=Ae^{-\beta{E}}=e^{-\psi-\beta{E}}
\end{equation}
Here $A$, $\beta$ and $\psi$ are constants. $E=H_0$ is the total energy of the system. In fact, according to Eq.(19), we have
\begin{equation}
{{\partial\rho}\over{\partial{x}_{i\sigma}}}=-\beta\rho{{\partial{H}_0}\over{\partial{x}_{i\sigma}}}~~~~~~~~~~~{{\partial\rho}\over{\partial{p}_{i\sigma}}}=-\beta\rho{{\partial{H}_0}\over{\partial{p}_{i\sigma}}}
\end{equation}
Put it into Eq.(18), we can prove that it is the solution of Eq.(18). Eq.(19) is just the canonical ensemble in the current statistical theory. It is known that this result has nothing to do with whether or not the retarded Lorentz forces and the damping radiation forces are introduced. Based on the current Liouville's equation and the definition of equilibrium states of isolated systems, we can get it directly. But unfortunately, this simple method is neglected by the current theory. Perhaps it is just this neglect, the hypothesis of equal probability became necessary and possible to exist.
\par
If the particles in the system are identical, and the force acted on each particle only relative to its coordinate, or the particles are independent each other without interactions between them, we have $E=\sum{E}_i$, $E_i=T_i+U_i$. Here $T_i$ is particle's kinetic energy, $U_i$ is potential energy, Eq.(19) becomes
\begin{equation}
\rho=Ae^{-\beta(E_1+\cdot\cdot\cdot{E}_i+\cdot\cdot\cdot+{E}_N)}=Ae^{-\beta{E}_1}\cdot\cdot\cdot{e}^{-\beta{E}_i}\cdot\cdot\cdot{e}^{-\beta{E}_N}
\end{equation}
The formula is just the distribution of near-independent subsystem. By the integral over while space and the normalization of the $\rho$ function, we get
\begin{equation}
\int^{+\infty}_{-\infty}\rho{d}\Omega=\int^{+\infty}_{-\infty}{A}^{1/{N}}{e}^{-\beta{E}_1}{d}\Omega_1\cdot\cdot\cdot\int^{+\infty}_{-\infty}{A}^{1/{N}}{e}^{-\beta{E}_i}{d}\Omega_i\cdot\cdot\cdot\int^{+\infty}_{-\infty}{A}^{1/{N}}{e}^{-\beta{E}_N}{d}\Omega_N=1
\end{equation}
Because the particles are identical, the forms of their energies $E_i$ are the same, so Eq.(22) can be written as
\begin{equation}
\int^{+\infty}_{-\infty}\rho{d}\Omega=(\int^{+\infty}_{-\infty}{A}^{1/{N}}{e}^{-\beta{E}_i}{d}\Omega_i)^N=1~~~~~~~or~~~~~~~~\int^{+\infty}_{-\infty}{A}^{1/{N}}{e}^{-\beta{E}_i}{d}\Omega_i=1
\end{equation}
Let $A^{1/N}=e^{-\alpha}$, the formula can be written as the form of sum
\begin{equation}
\sum_{i}{f}_i\Delta\Omega_i=\sum_{i}{e}^{-\alpha-\beta{E}_i}\Delta\Omega_i=1~~~~~~~~~~~~f_i\Delta\Omega_i=e^{-\alpha-\beta{E}_i}\Delta\Omega_i
\end{equation}
This is just the distribution law of the Maxwell-Boltzmann. or the distribution law of most probable value.
\par
If the particles are free without potential forces, we have $U_i=0$ and $E_{i}\rightarrow{m}v^2/{2}$. So the formula above becomes the Maxwell distribution law of velocities.
\begin{equation}
fd\Omega=Be^{-{{mv^2}\over{2KT}}}{d}xdydzdv_x{d}v_y{d}v_z
\end{equation}
Here $KT=1/\beta$, $B=e^{-\alpha}$. In this case, the distribution function has nothing to do with space coordinates, showing that the distribution of particles in space is uniform.
\par
When all particles have the same speed in system, according to Eq.(25), we have $f=$ constant and the system becomes micro-canonical ensemble. So micro-canonical ensemble represents the systems composed of free particles, in which particle's speeds are the same and the space distribution is even. This kind of systems, of course, has no significance in practice.
\par
On other hand, if there are two systems with the numbers of particles $N_1$ and $N_2$, energies $E_1$ and $E_2$ individually, we can also get the grand canonical ensemble as done in the current theory.
\par
So it is obvious that we can get all kinds of equivalent distributions from canonical ensemble, including micro-canonical ensemble. But we can't obtain canonical ensemble from micro-canonical ensemble. The situation is just opposite comparing with traditional theory. According to the current theory, canonical ensemble is deduced from micro-canonical ensemble, representing the equilibrium states of small systems to contact with big heat sources. So canonical ensemble describes the equilibrium states of non-isolated systems according to traditional statistical theory. But according to the Liouville's equation, canonical ensemble describes the equilibrium states of isolated systems. So micro-canonical ensemble contradicts the result of the Liouville's equation. On the other hand, micro-canonical ensemble is only one of equilibrium states. The other equilibrium states are not with equal probabilities. So it is improper to consider micro-canonical ensemble as the foundation of equilibrium state theory. For the systems only acted by conservative forces without considering retarded effects, the canonical ensemble is more foundational. For the general statistical systems, the revised Liouville's equation is fundamental. It can be seen from discussions above that it is more simple and rational to discuss the problems of equilibrium states from the Liouville's equation directly, no matter on physical concepts or mathematical calculation.
\par
3. When the non-conservative forces are considered and the system satisfies
\begin{equation}
\sum_{i\sigma}[{{\partial\rho}\over{\partial{x}_{i\sigma}}}{{p_{i\sigma}}\over{m_i}}+{{\partial\rho}\over{\partial{p}_{i\sigma}}}(F_{0i\sigma}+F_{i\sigma})]=0
\end{equation}
This is other equilibrium state. There are other equilibrium states.
\par
For non-isolated systems, we can definite the equilibrium states as the states that the forces acted on the systems and the ensemble probabilities functions do not change with time. For example, the equilibrium state of a system located in the gravitational field of earth. The distribution function can still be described by Eq.(19), but in this case total energy $E$ contains potential energy. In all theses equilibrium states, ensemble probability densities change with the coordinates of phase space, but do not change with time. 
\par
Therefore, it is clear that we can deal with all the problems of statistical physics well from the revised Liouville's equation directly. It is unnecessary for us to introduce some extra hypotheses such as the equal probability hypothesis. At present, chaos theory is also involved when we discuss the problems of foundation of statistical mechanics. Based on conservative interaction, thought chaos theory may be relative to the origin of random nature of statistical system or the hypothesis of ensemble, it has no effect on the foundational form of motion equation of statistical mechanics and has nothing to do with the origin of irreversibility in macro-systems. So it is impossible to us to solve the fundamental problems of statistical physics depending on chaos theory. In fact, it is enough for us in principle to solve all problems in statistical physics based on the revised Liouville's equation. By the method provided in the paper, the united description of equilibrium states and non-equilibrium states without using the concepts of ergoden, coarse graining and mixing currents and so on, which are easy to cause dispute.
\\
\\
{\large 5. The statistical distributions of non-equivalent states and BBGKY series equations }\\
\par
For non-equivalent states, by the direct integral of Eq.(15), we have 
\begin{equation}
\rho(t)=\exp[-\sum_{i\sigma}\int{{\partial{G}'_{i\sigma}}\over{\partial{p}_{i\sigma}}}dt]=\exp[-\sum^3_{\sigma=1}\sum^N_{i=1}\sum^N_{j\neq{1}}\int{{\partial{G}'_{ij\sigma}[\vec{r}_i,\vec{r}'_j,\vec{p}_i,\vec{p}'_j,\vec{a}'_j(\vec{r}_i,\vec{p}'_i,\vec{r}'_j,\vec{p}'_j)]}\over{\partial{p}_{i\sigma}}}dt]
\end{equation}
In order to let the integral is possible, the retarded quantities should be expressed by the non-retarded quantities or opposite. It is known that the relation between retarted time and particle's distance is
\begin{equation}
r_{ij}=c(t-t')=\sqrt{[x_i(t)-x'_j(t')]^2+[y_i(t)-y'_j(t')]^2+[z_i(t)-z'_j(t')]^2}
\end{equation}
Suppose the function relations $x_i=x_i(t)$ and $x'_j=x'_j(t')$ are known, it can be obtained from the formula above
\begin{equation}
t=f[x'_j(t'),y'_j(t'),z'_j(t'),t']~~~~~~~~~~~or~~~~~~~~~~t'=f'[x_i(t),y_i(t),z_i(t),t]
\end{equation}
From the relations, we have
\begin{equation}
dt={{\partial{f}}\over{\partial{t'}}}{d}t'+{{\partial{f}}\over{\partial{x}'_j}}{{dx'_j}\over{dt'}}{d}t'+{{\partial{f}}\over{\partial{y}'_j}}{{dy'_j}\over{dt'}}{d}t'+{{\partial{f}}\over{\partial{z}'_j}}{{dz'_j}\over{dt'}}{d}t'=({{\partial{f}}\over{\partial{t}}}+\vec{\nu}'_i\cdot\nabla'{f})dt'
\end{equation}
Put it into Eq.(27), we get the probability function expressing by retarded time
\begin{equation}
\rho(t)=\exp[-\sum^3_{\sigma=1}\sum^N_{i=1}\sum^N_{j\neq{i}}\int{{\partial{G}'_{ij\sigma}(\vec{r}_i,\vec{r}'_j,\vec{p}_i,\vec{p}'_j)}\over{\partial{p}_{i\sigma}}}({{\partial{f}}\over{\partial{t}'}}+\vec{\nu}'_i\cdot\nabla'{f})dt']
\end{equation}
On the other hand, by using relations $x_i=x_i(t)$, $y_i=y_i(t)$, $z_i=z_i(t)$ and Eq.(29), we have
$$r'_{ij}(t,t')=c[t-f'(x_i,y_i,z_i,t)]=r'_{ij}(t)$$
\begin{equation}
\vec{v}'_j=\vec{v}'_j[f'(x_i,y_i,z_i,t)]=\vec{\nu}'_j(t)~~~~~~~~~~~~~\vec{a}'_j=\vec{a}'_j[f'(x_i,y_i,z_i,t)]=\vec{a}'_j(t)
\end{equation}
Put them into Eq.(27), we get the probability function expressing by non-retarded time
\begin{equation}
\rho(t)=\exp[-\sum_{i\sigma}\int{{\partial{G}'_{i\sigma}}\over{\partial{p}_{i\sigma}}}dt]
\end{equation}
\par
When the particle's speed $v<<{c}$, the retarded distance $r'_{ij}(t')$ in the retarded time $t'$ can be replaced approximately by non-retarded distance $r_{ij}(t)$, i.e., we can let 
\begin{equation}
t'=t-r'_{ij}(t')/{c}\rightarrow{t}-r_{ij}(t)/{c}
\end{equation}
\begin{equation}
r'_{ij}(t,t')=\mid\vec{r}_i(t)-\vec{r}'_j(t')\mid=\mid\vec{r}_i(t)-\vec{r}'_j[t-r'_{ij}(t')/{c}]\mid\rightarrow\mid\vec{r}_i(t)-\vec{r}'_j[t-r_{ij}(t)/{c}]\mid=r'_{ij}(t)
\end{equation}
It is noted that $r'_{ij}(t)\neq{r}_{ij}(t)$, for $r'_{ij}(t)$ is the approximate retarded distance, but $r_{ij}(t)$ is not the retarded distance. In this case, we can develop retarded quantities into series in light of small quantity ${r}_{ij}/{c}$. By relation $\vec{v}_j=d\vec{r}_j/{d}t$, we get $^{(4)}$£º
$$\vec{r}'_{ij}(t,t')\simeq\vec{r}_i(t)-\vec{r}'_j(t-r_{ij}\prime{c})=\vec{r}_i(t)-\vec{r}_j(t)+{{r_{ij}(t)}\over{c}}\vec{v}_j(t)-{{r^2_{ij}(t)}\over{2c^2}}\vec{a}_j+{{r^3_{ij}(t)}\over{6c^3}}\dot{\vec{a}}_j+\cdot\cdot\cdot$$
\begin{equation}
=\vec{r}_{ij}(t)+{{r_{ij}(t)}\over{c}}\vec{v}_j(t)-{{r^2_{ij}(t)}\over{2c^2}}\vec{a}_j+{{r^3_{ij}(t)}\over{6c^3}}\dot{\vec{a}}_j+\cdot\cdot\cdot
\end{equation}
$$r'_{ij}(t')\simeq{r}_{ij}(t)\{1+{{\vec{r}_{ij}(t)\cdot\vec{\nu}_j(t)}\over{cr_{ij}(t)}}-{{\vec{r}_{ij}(t)\cdot\vec{a}_j(t)}\over{2c^2}}+{{\vec{r}_{ij}(t)\cdot\dot{\vec{a}}_j(t)}\over{6c^3}}{r}_{ij}(t)+\cdot\cdot\cdot\}$$
\begin{equation}
=r_{ij}\{1+{{v_{jn}}\over{c}}-{{a_{jn}{r}_{ij}}\over{2c^2}}+{{\dot{a}_{jn}{r}^2_{ij}}\over{6c^3}}+\cdot\cdot\cdot\}
\end{equation}
\begin{equation}
\vec{v}'_j(t')\simeq\vec{v}_j(t)-{{r_{ij}(t)}\over{c}}\vec{a}_j(t)+{{r^2_{ij}(t)}\over{2c^2}}\dot{\vec{a}}_j(t)+\cdot\cdot\cdot~~~~~~~~~~~~~\vec{a}'_j(t')\simeq\vec{a}_j(t)-{{r_{ij}(t)}\over{c}}\dot{\vec{a}}_j(t)+\cdot\cdot\cdot
\end{equation}
From Eq.(36) and (37), it is obvious that $\vec{r}'_{ij}(t')\neq\vec{r}_{ji}(t')$, $r'_{ij}(t')\neq{r}'_{ji}(t')$. The relations below are useful in later calculations.
\begin{equation}
{1\over{r'_{ij}}}\simeq{1\over{r_{ij}}}\{1-{{v_{jn}}\over{c}}+{{a_{jn}}\over{2c^2}}(1-{{2v_{jn}}\over{c}})r_{ij}-{{\dot{a}_{jn}}\over{6c^3}}(1-{{2v_{jn}}\over{c}})r^2_{ij}+\cdot\cdot\cdot\}
\end{equation}
\begin{equation}
{1\over{r'^2_{ij}}}\simeq{1\over{r^2_{ij}}}\{1-{{2v_{jn}}\over{c}}+{{3v^2_{jn}}\over{c^2}}+{{a_{jn}}\over{c^2}}(1-{{3v_{jn}}\over{c}})r_{ij}-{{\dot{a}_{jn}}\over{3c^3}}(1-{{3v_{jn}}\over{c}})r^2_{ij}+\cdot\cdot\cdot\}
\end{equation}
\begin{equation}
{1\over{r'^3_{ij}}}\simeq{1\over{r^2_{ij}}}\{1-{{3v_{jn}}\over{c}}+{{6v^2_{jn}}\over{c^2}}+{{3a_{jn}}\over{2c^2}}(1-{{5v_{jn}}\over{c}})r_{ij}-{{\dot{a}_{jn}}\over{2c^3}}(1-{{5v_{jn}}\over{c}})r^2_{ij}+\cdot\cdot\cdot\}
\end{equation}
\par
The BBGKY series equations are discussed now. The volume element of phase space of the i-particle is written as $d\Omega_i=\Pi_{\sigma=1}{d}x_{i\sigma}{d}p_{i\sigma}$ and the reduced distribution functions of ensemble probability densities are
\begin{equation}
f_s(x_{l\sigma},p_{l\sigma}\cdot\cdot\cdot{x}_{s\sigma},p_{s\sigma},t)=\int\rho(x_{l\sigma},p_{l\sigma}\cdot\cdot\cdot{x}_{n\sigma},p_{n\sigma},t)d\Omega_{S+1}\cdot\cdot\cdot{d}\Omega_N
\end{equation}
In the current theory, by considering the nature of identical particles, the BBGKY series equations are
$${{\partial{f}_S}\over{\partial{t}}}=-\sum^{3}_{\sigma=1}\sum^{S}_{i=1}({{p_{i\sigma}}\over{m_i}}{{\partial{f}_S}\over{\partial{x}_{i\sigma}}}+(F_{ei\sigma}{{\partial{f}_S}\over{\partial{p}_{i\sigma}}})-\sum^{3}_{\sigma=1}\sum^{S}_{i=1}\sum^S_{j\neq{i}}F_{ij\sigma}{{\partial{f}_S}\over{\partial{p}_{i\sigma}}}$$
\begin{equation}
-(N-S)\sum^3_{\sigma=1}\sum^S_{i=1}\int{F}_{iS+1\sigma}{{\partial{f}_{S+1}}\over{\partial{p}_{i\sigma}}}d\Omega_{s+1}
\end{equation}
The equations are equivalent to the Liouville's equation. Considering the fact that $F_{ei\sigma}$ only relative to the $i$ -particle's coordinates but $F'_{0i\sigma}$, $F'_{i\sigma}$ and $G'_{i\sigma}$ are relative to both the $i$ and $j$ particle's coordinates after the retarded Lorentz forces and radiation damping forces, we can get new BBGKY series equations in the same way 
$${{\partial{f}_S}\over{\partial{t}}}+\sum^{3}_{\sigma=1}\sum^{S}_{i=1}({{p_{i\sigma}}\over{m_i}}{{\partial{f}_S}\over{\partial{x}_{i\sigma}}}+F_{ei\sigma}{{\partial{f}_S}\over{\partial{p}_{i\sigma}}})+\sum^{3}_{\sigma=1}\sum^{S}_{i=1}\sum^S_{j\neq{i}}(F'_{0ij\sigma}+F'_{ij\sigma}){{\partial{f}_S}\over{\partial{p}_{i\sigma}}}$$
$$+(N-S)\sum^3_{\sigma=1}\sum^S_{i=1}\int(F'_{0iS+1\sigma}+F'_{iS+1\sigma}){{\partial{f}_{S+1}}\over{\partial{p}_{i\sigma}}}d\Omega_{s+1}$$
\begin{equation}
+\sum^3_{\sigma=1}\sum^S_{i=1}\sum^S_{j\neq{i}}{{\partial(G'_{ij\sigma}{f}_S)}\over{\partial{p}_{i\sigma}}}+(N-S)\sum^3_{\sigma=1}\sum^S_{i=1}\int{{\partial(G'_{iS+1\sigma}{f}_{S+1})}\over{\partial{p}_{i\sigma}}}d\Omega_{s+1}=0
\end{equation}
In the formula, $F'_{0iS+1,\sigma}$, $F'_{iS+1,\sigma}$ and $G'_{iS+1,\sigma}$ represents the $j=S+1$ items. Because $N-1\simeq{N}$, the first equation with $S=1$ is 
$${{\partial{f}_1}\over{\partial{t}}}+{{\vec{p}_1}\over{m}}\cdot\nabla_{\vec{r}_1}{f}_1+\vec{F}_{e1}\cdot\nabla_{\vec{p}_1}{f}_1+N\int\vec{F}'_{012}\cdot\nabla_{\vec{p}_1}{f}_2{d}^3\vec{r}_2{d}^3\vec{p}_2$$
\begin{equation}
=-N\int\vec{F}'_{12}\cdot\nabla_{\vec{p}_1}{f}_2{d}^3\vec{r}_2{d}^3\vec{p}_2-N\int\nabla_{\vec{p}_1}\cdot(\vec{G}'_{ij}{f}_2)d^3\vec{r}_2{d}^3\vec{p}_2
\end{equation}
In the formula, $\vec{F}'_{012}$ is the conservative retarded force and $\vec{F}'_{12}$ is the non-conservative retarded force. By using Eq.(36)----(41), $\vec{F}_{12}$ tcan be expressed by the non-retarded quantities as
\begin{equation}
\vec{F}_{0ij}={{q_i{q}_j\vec{r}'_{ij}}\over{r'^3_{ij}}}={{q_i{q}_j\vec{r}_{ij}}\over{r^3_{ij}}}+\vec{K}_{ij}=-\nabla_{\vec{r}_i}\cdot{U}(r_{ij})+\vec{K}_{ij}
\end{equation}
\begin{equation}
\vec{K}_{ij}={{q_i{q}_j\vec{r}_{ij}}\over{r^3_{ij}}}K_{ij1}+{{q_i{q}_j\vec{v}_j}\over{cr^2_{ij}}}K_{ij2}+{{q_i{q}_j\vec{a}_j}\over{c^2r_{ij}}}K_{ij3}+{{q_i{q}_j\dot{\vec{a}}_j}\over{c^3}}K_{ij4}
\end{equation}
\begin{equation}
K_{ij1}=-{{3v_{jn}}\over{c}}+{{6v^2_{jn}}\over{c^2}}+{{3a_{jn}}\over{2c^2}}(1-{{5v_{jn}}\over{c}})r_{ij}-{{\dot{a}_{jn}}\over{2c^3}}(1-{{5v_{jn}}\over{c}})r^2_{ij}+\cdot\cdot\cdot
\end{equation}
\begin{equation}
K_{ij2}=1-{{3v_{jn}}\over{c}}+{{3a_{jn}{r}_{ij}}\over{2c^2}}-{{\dot{a}_j{r}^2_{ij}}\over{2c^3}}
\end{equation}
\begin{equation}
K_{ij3}=-{1\over{2}}+{{3v_{jn}}\over{2c}}~~~~~~~K_{ij4}={1\over{6}}-{{v_{jn}}\over{2c}}
\end{equation}
It is obvious that after the retarded quantities are expressed by the non-retarded quantities, the conservative forces would become the non-conservative forces. In the same way, $\vec{G}'_{ij}$ can also be expressed by the non-retarded quantities. By using Eq.(46), Eq.(45) can be written as
$${{\partial{f}_1}\over{\partial{t}}}+{{\vec{p}_1}\over{m}}\cdot\nabla_{\vec{r}_1}{f}_1+\vec{F}_{e1}\cdot\nabla_{\vec{p}_1}{f}_1-N\int\nabla_{\vec{r}_1}{U}(r_{12})\cdot\nabla_{\vec{p}_1}{f}_2{d}^3\vec{r}_2{d}^3\vec{p}_2$$
\begin{equation}
=-N\int(\vec{K}_{12}+\vec{F}'_{12})\cdot\nabla_{\vec{p}_1}{f}_2{d}^3\vec{r}_2{d}^3\vec{p}_2-N\int\nabla_{\vec{p}_1}\cdot(\vec{G}'_{12}{f}_2)d^3\vec{r}_2{d}^3\vec{p}_2
\end{equation}
The left side of the equation is the result by the Liouville's equation. The right side is new by the paper's revision. Similarly, the second equation of the BBGKY series with $S=2$ can be written as
$${{\partial{f}_2}\over{\partial{t}}}+{{\vec{p}_1}\over{m}}\cdot\nabla_{\vec{r}_1}{f}_2+{{\vec{p}_2}\over{m}}\cdot\nabla_{\vec{r}_2}{f}_2+\vec{F}_{e1}\cdot\nabla_{\vec{p}_1}{f}_2+\vec{F}_{e2}\cdot\nabla_{\vec{p}_2}{f}_2-\nabla_{\vec{r}_1}{U}(r_{12})\cdot\nabla_{\vec{F}_1}{f}_2$$
$$-\nabla_{\vec{r}_2}{U}(r_{12})\cdot\nabla_{\vec{p}_2}{f}_2-N\int[\nabla_{\vec{r}_1}{U}(r_{13})\cdot\nabla_{\vec{p}_1}{f}_3+\nabla_{\vec{r}_2}{U}(r_{23})\cdot\nabla_{\vec{p}_2}{f}_3]d^3\vec{r}_3{d}^3\vec{p}_3$$
$$=-\nabla_{\vec{p}_1}\cdot(\vec{G}'_{12}{f}_2)-\nabla_{\vec{p}_2}\cdot(\vec{G}'_{21}{f}_2)+\vec{K}_{12}\cdot\nabla_{\vec{p}_1}{f}_2+\vec{K}_{21}\cdot\nabla_{\vec{p}_2}{f}_2$$
$$-N\int[(\vec{K}_{13}+\vec{F}'_{13})\cdot\nabla_{\vec{p}_1}{f}_3+(\vec{K}_{23}+\vec{F}'_{23})\cdot\nabla_{\vec{p}_2}{f}_3]{d}^3\vec{r}_3{d}^3\vec{p}_3$$
\begin{equation}
-N\int[\nabla_{\vec{p}_1}(\vec{G}'_{13}{f}_3)+\nabla_{\vec{p}_2}(\vec{G}'_{23}{f}_3)]d^3\vec{r}_3{d}^3\vec{p}_3
\end{equation}
The definition of $\vec{K}_{ij}$ can be seen in Eq.(47). We will discuss the motion equations of hydromechanics mechanics by using them below.
\\
\\
{\large 6. The motion equations of hydromechanics mechanics }\\
\par
How to reduce the motion equations of hydromechanics from statistical mechanics is still a unsolved problem now. Because the current statistical mechanics is based on the Liouville's equation, it is only suitable for the equivalent states of conservative systems or the ideal fluids. The dissipative phenomena of real fluids just as heat conduction and viscosity and so on can only be explained rationally by introducing non-conservative forces. Let's discuss this problem now. By writing the right items of Eq.(50) into the form of hydromechanics and adding them into the current results, we can get the results after the retarded Lorentz forces and radiation damping forces are introduced. According to the current theory, we definite the normalization of functions $f_1(\vec{r}_1.\vec{p}_1,t)$ and $f_2(\vec{r}_1.\vec{p}_1,\vec{r}_2,\vec{p}_2,t)$ as
\begin{equation}
V_0=\int{f}_1(\vec{r}_1,\vec{p}_1,t)d^3\vec{r}_1{d}^3\vec{p}_1
\end{equation}
\begin{equation}
V^2_0=\int{f}_2(\vec{r}_1,\vec{p}_1,\vec{r}_2,\vec{p}_2,t){d}^3\vec{r}_1{d}^3\vec{p}_1{d}^3\vec{r}_2{d}^3\vec{p}_2
\end{equation}
In the formula $\rho_0$ is macro-mass density, $\vec{V}$ is velocity of fluid, $u_k$ are dynamic energy density, $u_{v}$ is potential energy density $^{(5)}$£º
\begin{equation}
\rho_0(\vec{r}_1,t)={{mN}\over{V_0}}\int{f}_1(\vec{r}_1,\vec{p}_1,t)d^3\vec{p}_1~~~~~~~~\vec{V}(\vec{r}_1,t)={{N}\over{\rho_0{V}_0}}\int\vec{p}_1{f}_1(\vec{r}_1,\vec{p}_1,t)d^3\vec{p}_1
\end{equation}
\begin{equation}
u_k(\vec{r}_1,t)={{N}\over{2m\rho_0{V}_0}}\int(\vec{p}-m\vec{V})^2{f}_1(\vec{r}_1,\vec{p}_1,t)d^3\vec{p}_1
\end{equation}
\begin{equation}
u_{v}(\vec{r}_1,t)={1\over{2\rho_0}}({N\over{V_0}})^2\int{U}(\vec{r}_1,\vec{r}'_2){f}_2(\vec{r}_1,\vec{p}_2,\vec{r}'_2,\vec{p}'_2,t)d^3\vec{p}_1{d}^3\vec{r}_2{d}^3\vec{p}_2
\end{equation}
\par
The equilibrium equation of mass density or the continuity equation is discussed firstly. Producing Eq.(50) is produced by $mN/{V}_0$, then the integral about $d^3\vec{p}_1$ is carried out The right side of the equation can be written as
\begin{equation}
-{{N^2}\over{\rho_0{V}_0}}\int(\vec{K}_{12}+\vec{F}'_{12})\cdot\nabla_{\vec{p}_1}{f}_2{d}^3\vec{p}_1{d}^3\vec{r}_2{d}^3\vec{p}_2-{{N^2}\over{\rho_0{V}_0}}\int\nabla_{\vec{p}_1}\cdot(\vec{G}'_{12}{f}_2){d}^3\vec{p}_1{d}^3\vec{r}_2{d}^3\vec{p}_2
\end{equation}
According to Eq.(14), we have $\nabla_{\vec{p}_1}\cdot\vec{F}'_{12}=0$. Because $\vec{K}_{12}$ has nothing to do with $\vec{p}_1$, we also have $\nabla_{\vec{p}_1}\cdot\vec{K}_{12}=0$. By using the boundary conditions of probability density functions, Eq.(50) is zero. So the continuity equation of hydromechanics is the same as the current form with
\begin{equation}
{{\partial\rho_0}\over{\partial{t}}}+\nabla\cdot\rho_0\vec{V}=0
\end{equation}
\par
In order to get the motion equation of hydromechanics, we product Eq.(50) by $N\vec{p}_1/{V}_0$ and take the integral about $d^3\vec{p}_1$, then get from left side of Eq.(50)
\begin{equation}
\rho_0({{\partial\vec{V}}\over{\partial{t}}}+\vec{V}\cdot\nabla\vec{V})+\nabla\cdot(\vec{\vec{P}}_k+\vec{\vec{P}}_{\nu})=\rho_0\vec{\Gamma}_1
\end{equation}
In the formula, $\vec{\Gamma}_1$, $\vec{\vec{P}}_k$ and $\vec{\vec{P}}_v$ are the average external force, kinetic energy and potential energy tensors of unit masses individually $^{(5)}$£º
\begin{equation}
\vec{\Gamma}_1={N\over{\rho_0{V}_0}}\int{f}_1\vec{F}_{e1}{d}^3\vec{p}_1~~~~~~~~~\vec{\vec{P}}_k={N\over{V_0}}\int{f}_1{{(\vec{p}_1-m\vec{V})(\vec{p}_1-m\vec{V})}\over{m}}{d}^3\vec{p}_1
\end{equation}
\begin{equation}
\vec{\vec{p}}_{\nu}=-{1\over{2}}({N\over{V_0}})^2\int^1_0{d}\lambda\int{{\vec{r}''_{12}\vec{r}''_{12}}\over{r''_{12}}}{{dU(r''_{12})}\over{dr''_{12}}}{d}^3{r}''_{12}{d}^3\vec{p}_1{d}^3\vec{p}_2
\end{equation}
For the right side of Eq.(50), because $\nabla_{\vec{p}_1}\cdot(\vec{K}_{12}+\vec{F}'_{12})=0$, we get
$$-{N^2\over{V_0}}\int\vec{p}_1(\vec{K}_{12}+\vec{F}'_{12})\cdot\nabla_{\vec{p}_1}{f}_2{d}^3\vec{p}_1{d}^3\vec{r}_2{d}^3\vec{p}_2-{N\over{V_0}}\int\vec{p}_1\nabla_{\vec{p}_1}\cdot(\vec{G}'_{12}{f}_2){d}^3\vec{p}_1{d}^3\vec{r}_2{d}^3\vec{p}_2$$
$$={N^2\over{V_0}}\int{f}_2(\vec{K}_{12}+\vec{F}'_{12})\cdot\nabla_{\vec{p}_1}\vec{p}_1{d}^3\vec{p}_1{d}^3\vec{r}_2{d}^3\vec{p}_2+{N^2\over{V_0}}\int{f}_2\vec{G}'_{12}\cdot\nabla_{\vec{p}_1}\vec{p}_1{d}^3\vec{p}_1{d}^3\vec{r}_2{d}^3\vec{p}_2$$
\begin{equation}
={N^2\over{V_0}}\int{f}_2(\vec{K}_{12}+\vec{F}'_{12})d^3\vec{p}_1{d}^3\vec{r}_2{d}^3\vec{p}_2+{{N^2}\over{V_0}}\int{f}_2\vec{G}'_{12}{d}^3\vec{p}_1{d}^3\vec{r}_2{d}^3\vec{p}_2
\end{equation}
On the other hand, by using Eq.(36)---(41), Eq.(5) can be expressed by the non-retarded quantities
\begin{equation}
\vec{F}'_{ij}={{q_i{q}_j\vec{r}_{ij}}\over{r^3_{ij}}}{Q}_{ij1}+{{q_i{q}_j\vec{v}_{ij}}\over{cr^2_{ij}}}{Q}_{ij2}+{{q_i{q_j}\vec{a}_{ij}}\over{c^2{r}_ij}}{Q}_{ij3}+{{q_i{q}_j\dot{\vec{a}}_{ij}}\over{c^3}}{Q}_{ij4}
\end{equation}
Let $R_{ijk}=K_{ijk}+Q_{ijk}$, we can write $\vec{K}_{12}+\vec{F}'_{12}$ as
\begin{equation}
\vec{K}_{12}+\vec{F}'_{12}={{q_1{q}_2\vec{r}_{12}}\over{r^3_{12}}}{R}_{121}+{{q_1{q}_2\vec{v}_2}\over{cr^2_{12}}}{R}_{122}+{{q_1{q}_2\vec{a}_2}\over{c^2{r}_{12}}}{R}_{123}+{{q_1{q}_2\dot{\vec{a}}_2}\over{c^3}}{R}_{124}
\end{equation}
In the formula, $R_{ijk}$ does not contain retarded quantities again. By the relation $q_1{q}_2\vec{r}^3_{12}=-\nabla_{\vec{r}_1}{U}(r_{12})$, we have
$$-{N^2\over{V_0}}\int(\vec{K}_{12}+\vec{F}'_{12}){f}_2{d}^3\vec{p}_1{d}^3\vec{r}_2{d}^3\vec{p}_2$$
$$={{N^2}\over{V_0}}\int\nabla_{\vec{r}_1}{U}(r_{12})R_{121}{f}_2{d}^3\vec{p}_1{d}^3\vec{r}_2{d}^3\vec{p}_2-{{N^2}\over{V_0}}\int{{q_1{q_2}\vec{v}_2}\over{cr^2_{12}}}R_{122}{f}_2{d}^3\vec{p}_1{d}^3\vec{r}_2{d}^3\vec{p}_2$$
\begin{equation}
-{{N^2}\over{V_0}}\int{{q_1{q}_2\vec{a}_2}\over{cr^2_{12}}}R_{123}{f}_2{d}^3\vec{p}_1{d}^3\vec{r}_2{d}^3\vec{p}_2-{{N^2}\over{V_0}}\int{{q_1{q_2}\vec{a}_2}\over{cr^2_{12}}}R_{124}{f}_2{d}^3\vec{p}_1{d}^3\vec{r}_2{d}^3\vec{p}_2
\end{equation}
By using formula $^{(5)}$ 
\begin{equation}
\int{f}(\vec{r}_1,\vec{r}_2)\nabla_{r_1}{U}(r_{12}){d}^3\vec{r}_2=\nabla_{r_1}\cdot\{-{1\over{2}}\int^1_0{d}\lambda\int{d}^3\vec{r}''_{12}{{\vec{r}''_{12}\vec{r}''_{12}}\over{r''_{12}}}{{dU(r''_{12})}\over{dr''_{12}}}{f}(\vec{r}_1+(1-\lambda)\vec{r}''_{12},\vec{r}_1-\lambda\vec{r}''_{12})\}
\end{equation}
we have
\begin{equation}
{{N^2}\over{V_0}}\int\nabla_{\vec{r}_1}{U}(r_{12}){R}_{121}{f}_2{d}^3\vec{p}_1{d}^3\vec{r}_2{d}^3\vec{p}_2=-\nabla_{\vec{r}_1}\cdot\vec{\vec{P}}_s
\end{equation}
$$\vec{\vec{p}}_s={{N^2}\over{2V_0}}\int^1_0{d}\lambda\int{{\vec{r}''_{12}\vec{r}''_{12}}\over{r''_{12}}}{{dU(r''_{12})}\over{dr''_{12}}}{R}_{121}[\vec{r}_1+(1-\lambda)\vec{r}''_{12},\vec{p}_1,\vec{r}_1-\lambda\vec{r}''_{12},\vec{p}_2,t]$$
\begin{equation}
\times{f}_2[\vec{r}_1+(1-\lambda)\vec{r}''_{12},\vec{p}_1,\vec{r}_1-\lambda\vec{r}''_{12},\vec{p}_2,t]{d}^3\vec{r}''_{12}{d}^3\vec{p}_1{d}^3\vec{p}_2
\end{equation}
We call $\vec{\vec{P}}_s$ as dissipative energy tensor. Let
\begin{equation}
\vec{\Gamma}_2=-{{q_1{q}_2{N}^2}\over{c\rho_0{V}_0}}\int({{\vec{v}_2}\over{r^2_{12}}}{R}_{122}+{{\vec{a}_2}\over{cr_{12}}}{R}_{123}+{{\dot{\vec{a}}_2}\over{c^2}}{R}_{124}){f}_2{d}^3\vec{p}_1{d}^3\vec{r}_2{d}^3\vec{p}_2
\end{equation}
represent the average dissipative force acted on unite mass and 
\begin{equation}
\vec{\Gamma}_3={{N^2}\over{\rho_0{V}_0}}\int{f}_2\vec{G}'_{12}{d}^3\vec{p}_1{d}^3\vec{r}_2{d}^3\vec{p}_2
\end{equation}
represent the average radiation damping force of unite mass, the motion equation of hydromechanics is
\begin{equation}
\rho_0({{\partial\vec{V}}\over{\partial{t}}}+\vec{V}\cdot\nabla\vec{V})+\nabla\cdot(\vec{\vec{P}}_k+\vec{\vec{P}}_v+\vec{\vec{P}}_s)=\rho_0(\vec{\Gamma}_1+\vec{\Gamma}_2+\vec{\Gamma}_3)
\end{equation}
Producing Eq.(50) by $N(\vec{p}_1-m\vec{V})^2/{2}mV_0$ and taking the integral about $d^3\vec{p}_1$, we can get the equivalent equation of kinetic energy. Form the left side of the equation, we have
\begin{equation}
{{\partial(\rho_0{u}_k)}\over{\partial{t}}}+\nabla\cdot(\rho_0{u}_k\vec{V}+\vec{J}_k)=-\vec{\vec{P}}_k:\nabla\vec{V}
\end{equation}
Here $\vec{J}_k=\vec{J}_{k1}+\vec{j}_{k2}$ with
\begin{equation}
\vec{J}_{k1}={N\over{V_0}}\int{{\vec{p}-m\vec{V}}\over{m}}{{(\vec{p}_1-m\vec{V})^2}\over{2m}}{f}_1{d}^3\vec{p}_1
\end{equation}
\begin{equation}
\vec{J}_{k2}=\int^1_0{d}\lambda{{\vec{r}''_{12}\vec{r}''_{12}\cdot(\vec{p}_1-m\vec{V})}\over{2mr''_{12}}}{{dU(r''_{12})}\over{dr''_{12}}}{f}_2(\vec{r}_1+(1-\lambda)\vec{r}''_{12},\vec{p}_2,\vec{r}_1-\lambda\vec{r}''_{12},\vec{p}_2,t)d^3\vec{r}''_{12}{d}^3\vec{p}_1{d}^3\vec{p}_2
\end{equation}
Form the left side of the equation, we get
$$-{{N^2}\over{2mV_0}}\int(\vec{p}_1-m\vec{V})^2\nabla_{\vec{p}_1}\cdot[(\vec{K}+\vec{F}'_{12})f_2]d^3\vec{p}_1{d}^3\vec{r}_2{d}^3\vec{p}_2$$
$$-{{N}\over{2mV_0}}\int(\vec{p}_1-m\vec{V})^2\nabla_{\vec{p}_1}\cdot(\vec{G}'_{12}{f}_2)d^3\vec{p}_1{d}^3\vec{r}_2{d}^3\vec{p}_2$$
$$={{N^2}\over{mV_0}}\int{f}_2(\vec{K}_{12}+\vec{F}'_{12})\cdot(\vec{p}_1-m\vec{V})d^3\vec{p}_1{d}^3\vec{r}_2{d}^3\vec{p}_2$$
\begin{equation}
+{{N^2}\over{mV_0}}\int{f}_2\vec{G}'_{12}\cdot(\vec{p}_1-m\vec{V})d^3\vec{p}_1{d}^3\vec{r}_2{d}^3\vec{p}_2
\end{equation}
By using Eq.(65), the first item of the formula above can be written as
\begin{equation}
{{q_1{q}_2{N}^2}\over{mV_0}}\int{f}_2({{\vec{r}_{12}}\over{r^3_{12}}}{R}_{121}+{{\vec{\nu}_2}\over{cr^2_{12}}}{R}_{122}+{{\vec{a}_2}\over{c^2{r}_{12}}}{R}_{123}+{{\dot{\vec{a}_2}}\over{c^3}}{R}_{124})\cdot(\vec{p}_1-m\vec{V}){d}^3\vec{p}_1{d}^3\vec{r}_2{d}^3\vec{p}_2
\end{equation}
By using Eq.(67) again, the first item of the formula above can be written as
$$\nabla\cdot\vec{J}_{k3}={{N^2}\over{mV_0}}\int{f}_2{R}_{121}{{q_1{q}_2\vec{r}_{12}}\over{r^3_{12}}}\cdot(\vec{p}_1-m\vec{V})d^3\vec{p}_1{d}^3\vec{r}_2{d}^3\vec{p}_2$$
$$\vec{J}_{k3}={{N^2}\over{2mV_0}}\int^1_0{d}\lambda\int{{\vec{r}''_{12}\vec{r}''_{12}\cdot(\vec{p}_1-m\vec{V})}\over{r''_{12}}}{{dU(r''_{12})}\over{dr''_{12}}}{R}_{121}[\vec{r}_1+(1-\lambda)\vec{r}''_{12},\vec{p}_1,\vec{r}_1-\lambda\vec{r}''_{12},\vec{p}_2,t]$$
\begin{equation}
\times{f}_2[\vec{r}_1+(1-\lambda)\vec{r}''_{12},\vec{p}_1,\vec{r}_1-\lambda\vec{r}''_{12},\vec{p}_2,t]d^3\vec{r}''_{12}{d}^3\vec{p}_1{d}^3\vec{p}_2
\end{equation}
$\vec{J}_{k3}$ can be called as the dissipative fluid of kinetic energy of unite mass. The second, third and forth items in Eq.(77) and the last item in Eq.(76) can be put together and written as
\begin{equation}
\rho_0\sigma_k={{q_1{q}_2{N}^2}\over{cmV_0}}\int{f}_2({{\vec{\nu}_2}\over{r^2_{12}}}Q_{122}+{{\vec{a}_2}\over{cr_{12}}}Q_{123}+{{\dot{\vec{a}}_2}\over{c^2}}Q_{124}+{{cm}\over{q_1{q_2}}}\vec{G}'_{12})\cdot(\vec{p}_1-m\vec{V})d^3\vec{p}_1{d}^3\vec{r}_2{d}^3\vec{p}_2
\end{equation}
We can call $\sigma_k$ as the dissipative kinetic energy generation of unite mass. In the current theory, there is no this item. It appears only when the retarded Lorentz forces and damping radiation forces are introduced. Let $\vec{J}_k=\vec{J}_{k1}+\vec{J}_{k2}+\vec{J}_{k3}$, thus Eq (73) can be written as
\begin{equation}
{{\partial(\rho_0{u}_k)}\over{\partial{t}}}+\nabla\cdot(\rho_0{u}_k\vec{V}+\vec{J}_k)=-\vec{\vec{p}}_k:\nabla\vec{V}+\rho_0\sigma_k
\end{equation}
\par
Producing Eq.(51) by $N^2{U}(r_{12})/{V}^2_0$ and taking the integral about $d^3\vec{p}_1{d}^3\vec{r}_2{d}^3\vec{p}_2$, the equivalent equation of potential energy can be obtained. Form the left side of the equation, we can get 
\begin{equation}
{{\partial(\rho_0{u}_{\nu})}\over{\partial{t}}}+\nabla\cdot(\rho_0{u}_{\nu}\vec{V}+\vec{J}_{\nu})=-\vec{\vec{p}}_{\nu}:\nabla\vec{V}
\end{equation}
Here $\vec{J}_v=\vec{J}_{v1}+\vec{J}_{v2}$ with
\begin{equation}
\vec{J}_{\nu{1}}={1\over{2}}({N\over{V_0}})^2\int{{\vec{p}-m\vec{V}}\over{m}}U(r_{ij})f_2{d}^3\vec{p}_1{d}^3\vec{r}_2{d}^3\vec{p}_2
\end{equation}
\begin{equation}
\vec{J}_{\nu{2}}=-\int^1_0{d}\lambda{{\vec{r}''_{12}\vec{r}''_{12}\cdot(\vec{p}_1-\vec{p}_2)}\over{4mr''_{12}}}{{dU(r''_{12})}\over{dr''_{12}}}{f}_2(\vec{r}_1+(1-\lambda)\vec{r}''_{12},\vec{p}_2,\vec{r}_1-\lambda\vec{r}''_{12},\vec{p}_2,t){d}^3\vec{r}''_{12}{d}^3\vec{p}_1{d}^3\vec{p}_2
\end{equation}
The right side of the equation is
$${{N^2}\over{V^2_0}}\int{U}(r_{12})\vec{K}_{12}\cdot\nabla_{\vec{p}_1}{f}_2{d}^3\vec{p}_1{d}^3\vec{r}_2{d}^3\vec{p}_2+{{N^2}\over{V^2_0}}\int{U}(r_{12})\vec{K}_{21}\cdot\nabla_{\vec{p}_2}{f}_2{d}^3\vec{p}_1{d}^3\vec{r}_2{d}^3\vec{p}_2$$
$$-{{N^3}\over{V^2_0}}\int{U}(r_{12})[(\vec{K}_{13}+\vec{F}'_{13})\cdot\nabla_{\vec{p}_1}{f}_3+(\vec{K}_{23}+\vec{F}'_{23})\cdot\nabla_{\vec{p}_2}{f}_3]{d}^3\vec{p}_1{d}^3\vec{r}_2{d}^3\vec{p}_2{d}^3\vec{r}_3{d}^3\vec{p}_3$$
$$-{{N^2}\over{V^2_0}}\int{U}(r_{12})\nabla_{\vec{p}_1}\cdot(\vec{G}'_{12}{f}_2){d}^3\vec{p}_1{d}^3\vec{r}_2{d}^3\vec{p}_2-{{N^2}\over{V^2_0}}\int{U}(r_{12})\nabla_{\vec{p}_2}\cdot(\vec{G}'_{21}{f}_2){d}^3\vec{p}_1{d}^3\vec{r}_2{d}^3\vec{p}_2$$
\begin{equation}
-N\int{U}(r_{12})[\nabla_{\vec{p}_1}\cdot(\vec{G}'_{13}{f}_3)+\nabla_{\vec{p}_2}\cdot(\vec{G'_{23}{f}_3})]d^3\vec{p}_1{d}^3\vec{r}_2{d}^3\vec{p}_2{d}^3\vec{r}_3{d}^3\vec{p}_3
\end{equation}
As mentioned before, we have $\nabla_{\vec{p}_i}\cdot\vec{K}_{ij}=0$, $\nabla_{\vec{p}_i}\cdot\vec{K}'_{ij}=0$. By considering the boundary condition, all integrals are zero. So the potential energy items have no reversion after retarded Lorentz forces and the radiation damping forces are introduced. 
\par
We have to consider the contributions of non-conservative dissipative energies. It is known in electromagnetic theory that when the common coordinates and momentums are used to describe particle's motion, interaction energy does not contain magnetic field's action. The retarded interaction between the $i$ -particle and the $j$ -particle is 
\begin{equation}
U'_{ij}(r'_{ij},\vec{p}'_j)={{q_i{q}_j}\over{r'_{ij}(1-\nu'_{jn}\prime{c})}}={{q_i{q}_j}\over{r'_{ij}}}(1+{{\nu'_{jn}}\over{c}}+{{\nu'^2_{jn}}\over{c^2}})
\end{equation}
$U'_{ij}$ is asymmetry about the indexes $i$ and $j$ after the retarded effect is considered, so the non-conservative and retarded total interaction can be written as 
$$U'(r'_{ij},\vec{p}'_j)={1\over{2}}\sum^N_{i=1}\sum^N_{j\neq{1}}{{q_i{q}_j}\over{r'_{ij}}}({{\nu'_{jn}}\over{c}}+{{\nu'^2}\over{c^2}})={1\over{2}}\sum^N_{i=1}\sum^N_{j\neq{1}}{{q_i{q}_j}\over{r_{ij}}}$$
\begin{equation}
\times\{{{v_{jn}}\over{c}}-{{a_{ij}{r}_{ij}}\over{c^2}}(1-{{v_{jn}}\over{c}})+{{\dot{a}_{jn}{r}^2_{ij}}\over{2c^3}}(1+{{4v_{jn}}\over{3c}})-{{3\vec{v}_j\cdot\vec{a}_j{r}_{ij}}\over{2c^3}}+{{2\vec{v}_j\cdot\dot{\vec{a}}_j{r}^2_{ij}}\over{3c^4}}\}
\end{equation}
Producing the second equation of the BBGKY series equation by $N^2{U}'(r'_{12},\vec{p}_1,\vec{p}'_2)/{2}V^2_0$ and taking the integral about $d^3\vec{p}_1{d}^3\vec{r}_2{d}^3\vec{p}_2$, we get
$${{\partial}\over{\partial{t}}}{{N^2}\over{2V^2_0}}\int{U'}f_2{d}^3\vec{p}_1{d}^3\vec{r}_2{d}^3\vec{p}_2+{{N^2}\over{2mV^2_0}}\int{U'}(\vec{p}_1\cdot\nabla_{\vec{r}_q}+\vec{p}_2\cdot\nabla_{\vec{r}_2}){f}_2{d}^3\vec{p}_1{d}^3\vec{r}_2{d}^3\vec{p}_2$$
$$+{{N^2}\over{2V^2_0}}\int{U'}(\vec{F}_{e1}\cdot\nabla_{\vec{p}_1}+\vec{F}_{e2}\cdot\nabla_{\vec{p}_2})f_2{d}^3\vec{p}_1{d}^3\vec{r}_2{d}^3\vec{r}_2{d}^3\vec{p}_2$$
$$-{{N^2}\over{2V^2_0}}\int{U'}[\nabla_{\vec{r}_1}{U}(r_{12})\cdot\nabla_{\vec{p}_1}+\nabla_{\vec{r}_2}{U}(r_{12})\cdot\nabla_{\vec{p}_2}]f_2{d}^3\vec{p}_1{d}^3\vec{r}_2{d}^3\vec{p}_2$$
$$-{{N^3}\over{2V^2_0}}\int{U'}[\nabla_{\vec{r}_1}{U}(r_{13})\cdot\nabla_{\vec{p}_1}+\nabla_{\vec{r}_2}{U}(r_{23})\cdot\nabla_{\vec{p}_2}]f_3{d}^3\vec{p}_1{d}^3\vec{r}_2{d}^3\vec{p}_2{d^3}\vec{r}_3{d}^3\vec{p}_3$$
$$=-{{N^2}\over{2V^2_0}}\int{U'}[\nabla_{\vec{p}_1}\cdot(\vec{G}'_{12}{f}_2)+\nabla_{\vec{p}_2}\cdot(\vec{G}'_{21}{f}_2)]{d}^3\vec{p}_1{d}^3\vec{r}_2{d}^3\vec{p}_2$$
$$+{{N^2}\over{2V^2_0}}\int{U'}(\vec{K}_{12}\cdot\nabla_{\vec{p}_1}+\vec{K}_{21}\cdot\nabla_{\vec{p}_2})f_2{d}^3\vec{p}_1{d}^3\vec{r}_2{d}^3\vec{p}_2$$
$$-{{N^3}\over{2V^2_0}}\int{U'}[(\vec{K}_{13}+\vec{F}'_{13})\cdot\nabla_{\vec{p}_1}f_3+(\vec{K}_{23}+\vec{F}'_{23})\cdot\nabla_{\vec{p}_2}f_3]{d}^3\vec{p}_1{d}^3\vec{r}_2{d}^3\vec{p}_2{d}^3\vec{r}_3{d}^3\vec{p}_3$$
\begin{equation}
-{{N^3}\over{2V^2_0}}\int{U'}[\nabla_{\vec{p}_1}\cdot(\vec{G}'_{13}{f}_3)+\nabla_{\vec{p}_2}\cdot(\vec{G}'_{23}{f}_3)]d^3\vec{p}_1{d}^3\vec{r}_2{d}^3\vec{p}_2{d}^3\vec{r}_3{d}^3\vec{p}_3
\end{equation}
The calculation is verbose and we only give out the result of the first item in the right side of Eq.(86)
\begin{equation}
U_1={{q_1{q}_2}\over{2c}}({{v_{2n}}\over{r_12}}+{{v_{1n}}\over{r_{21}}})={{q_1{q}_2}\over{2c}}({{\vec{r}_{12}\cdot\vec{v}_1}\over{r^2_{12}}}+{{\vec{r}_{21}\cdot\vec{v}_1}\over{r^2_{21}}})={{q_1{q}_2}\over{2cm}}{{\vec{r}_{12}\cdot(\vec{p}_2-\vec{p}_1)}\over{r^2_{12}}}
\end{equation}
Similar to Eq.(56), we definite
\begin{equation}
u_s={1\over{2\rho_0}}{{N^2}\over{V^2_0}}\int{U}_1{f}_2{d}^3\vec{p}_1{d}^3\vec{r}_2{d}^3\vec{p}_2
\end{equation}
\begin{equation}
\vec{J}_{s1}={1\over{2}}{{N^2}\over{V^2_0}}\int{{{\vec{p}_1-m\vec{V}}\over{m}}}{U}_1{f}_2{d}^3\vec{p}_1{d}^3\vec{r}_2{d}^3\vec{p}_2
\end{equation}
as the energy density and energy fluid density of non-conservative interactions of unite mass and have
\begin{equation}
\rho_0{u}_s\vec{V}={1\over{2}}({N\over{V_0}})^2\int\vec{V}{U}_1{f}_2{d}^3\vec{p}_1{d}^3\vec{r}_2{d}^3\vec{p}_2
\end{equation}
$$\nabla_{\vec{r}_i}\cdot(\rho_0{u}_s\vec{V}+\vec{J}_{s1})=\nabla_{\vec{r}_1}\cdot\{{1\over{2}}{{N^2}\over{V^2_0}}\int{{\vec{p}_1}\over{m}}{U}_1{f}_2{d}^3\vec{p}_1{d}^3\vec{r}_2{d}^3\vec{p}_2\}$$
\begin{equation}
={1\over{2}}{{N^2}\over{V^2_0}}\int({U}_1{{\vec{p}_1}\over{m}}\cdot\nabla_{\vec{r}_1}{f}_2+{f}_2{{\vec{p}_1}\over{m}}\cdot\nabla_{\vec{r}_1}{U}_1){d}^3\vec{p}_1{d}^3\vec{r}_2{d}^3\vec{p}_2
\end{equation}
By considering the boundary condition, we can get
\begin{equation}
{{N^2}\over{2mV^2_0}}\int{U}_1\vec{p}_2\cdot\nabla_{\vec{r}_2}{f}_2{d}^3\vec{p}_1{d}^3\vec{r}_2{d}^3\vec{p}_2=-{{N^2}\over{2mV^2_0}}\int{f}_2\vec{p}_2\cdot\nabla_{\vec{r}_2}U_1{d}^3\vec{p}_1{d}^3\vec{r}_2{d}^3\vec{p}_2
\end{equation}
Therefore, we have
$${{N^2}\over{2mV^2_0}}\int{U}_1(\vec{p}_1\cdot\nabla_{\vec{r}_1}+\vec{p}_2\cdot\nabla_{\vec{r}_2}){f}_2{d}^3\vec{p}_1{d}^3\vec{r}_2{d}^3\vec{p}_2$$
\begin{equation}
=\nabla_{\vec{r}_1}\cdot(\rho_0{u}_s\vec{V}+\vec{J}_{s1})-{1\over{2}}{{N^2}\over{V^2_0}}\int{f}_2{{\vec{p}_1-\vec{p}_2}\over{m}}\cdot\nabla_{\vec{r}_1}U_1{d}^3\vec{p}_1{d}^3\vec{r}_2{d}^3\vec{p}_2
\end{equation}
For Eq.(88), because
\begin{equation}
{{\partial}\over{\partial{x}_1}}{{(x_1-x_2)(p_{2x}-p_{1x})}\over{r^2_{12}}}={{p_{2x}-p_{1x}}\over{r^2_{12}}}-{{(x_1-x_2)^2(p_{2x}-p_{1x})}\over{r^4_{12}}}
\end{equation}
by considering symmetry, we let
\begin{equation}
\nabla_{\vec{r}_1}{{\vec{r}_{12}\cdot(\vec{p}_2-\vec{p}_1)}\over{r^2_{12}}}\simeq{{\vec{p}_2-\vec{p}_1}\over{r^2_{12}}}-{{\vec{p}_2-\vec{p}_1}\over{3r^2_{12}}}=-{{2(\vec{p}_1-\vec{p}_2)}\over{3r^2_{12}}}
\end{equation}
The second item in the left sideof Eq.(94) can be written as
$$-{{q_1{q}_2{N}^2}\over{4cmV^2_0}}\int{f}_2{{(\vec{p}_1-\vec{p}_2)}\over{m}}\nabla_{\vec{r}_1}{{\vec{r}_{12}\cdot(\vec{p}_2-\vec{p}_1)}\over{r^2_{12}}}d^3\vec{p}_1{d}^3\vec{r}_2{d}^3\vec{p}_2$$
\begin{equation}
={{q_1{q}_2{N}^2}\over{6cm^2V^2_0}}\int{f}_2{{(\vec{p}_1-\vec{p}_2)^2}\over{r^2_{12}}}{d}^3\vec{p}_1{d}^3\vec{r}_2{d}^3\vec{p}_2=-\rho_0\sigma_s
\end{equation}
$\sigma_{s1}$ can be called as the production of dissipative energy. Because
\begin{equation}
\nabla_{\vec{p}_1}{U}_1={{q_1{q}_2}\over{2cm}}\nabla_{\vec{p}_1}{{\vec{r}_{12}\cdot(\vec{p}_2-\vec{p}_1)}\over{r^2_{12}}}=-{{q_1{q}_2}\over{2cm}}{{\vec{r}_{12}}\over{r^2_{12}}}~~~~~~~~~~~~~\nabla_{\vec{p}_2}U_1={{q_1{q}_2}\over{2cm}}{{\vec{r}_{12}}\over{r^2_{12}}}
\end{equation}
The third item in the left side of Eq.(88) can be written as
$$-{{N^2}\over{2V^2_0}}\int{f}_2(\vec{F}_{e1}\cdot\nabla_{\vec{p}_1}+\vec{F}_{e2}\cdot\nabla_{\vec{p}_2})U_1{d}^3\vec{p}_1{d}^3\vec{r}_2{d}^3\vec{p}_2={{q_1{q}_2{N}^2}\over{4cmV^2_0}}\int{f}_2{{\vec{r}_{12}\cdot(\vec{F}_{e1}-\vec{F}_{e2})}\over{r^2_{12}}}{d}^3\vec{p}_1{d}^3\vec{r}_2{d}^3\vec{p}_2$$
\begin{equation}
=-{{N^2}\over{4cmV^2_0}}\int{f}_2\nabla_{\vec{r}_1}{U}(r_{12})\cdot{r}_{12}(\vec{F}_{e1}-\vec{F}_{32}){d}^3\vec{p}_1{d}^3\vec{r}_2{d}^3\vec{p}_2=\nabla\cdot\vec{J}_{s2}
\end{equation}
It can be proved that Eq.(67) still hold when $f$ is a vector, so let $\vec{f}=f(\vec{F}_{e1}-\vec{F}_{e2})$, we have
$$\vec{J}_{s2}={{N^2}\over{8cmV^2_0}}\int^{1}_{0}{d}\lambda\int{{\vec{r}''_{12}\vec{r}''_{12}\cdot{r}''_{12}[\vec{F}_{e1}(\vec{r}_1+(1-\lambda)\vec{r}''_{12},t)-\vec{F}_{e2}(\vec{r}_1-\lambda\vec{r}''_{12},t)]}\over{r''_{12}}}{{dU(r''_{12})}\over{dr''_{12}}}$$
$$\times{f}_2(\vec{r}_1+(1-\lambda)\vec{r}''_{12},\vec{p}_2,\vec{r}_1-\lambda\vec{r}''_{12},\vec{p}_2,t)d^3r''_{12}{d}^3\vec{p}_1{d}^3\vec{p}_2$$
$$={{N^2}\over{8cmV^2_0}}\int^1_0{d}\lambda\int\vec{r}''_{12}\vec{r}''_{12}\cdot[\vec{F}_{e1}(\vec{r}_1+(1-\lambda)\vec{r}''_{12},t)-\vec{F}_{e2}(\vec{r}_1-\lambda\vec{r}''_{12},t)]{{dU(r''_{12})}\over{dr''_{12}}}$$
\begin{equation}
\times{f}_2(\vec{r}_1+(1-\lambda)\vec{r}''_{12},\vec{p}_2,\vec{r}_1-\lambda\vec{r}'_{12},\vec{p}_2,t)d^3r''_{12}{d}^3\vec{p}_1{d}^3\vec{p}_2
\end{equation}
By using Eq.(98) and (67) as well as relation $\nabla_{\vec{r}_1}{U}(r_{12})=-\nabla_{\vec{r}_2}{U}(r_{12})$, the fourth item in the left side of Eq.(87) is
$$-{{N^2}\over{2V^2_0}}\int{U}_1[\nabla_{\vec{r}_1}{U}(r_{12})\cdot\nabla_{\vec{p}_1}+\nabla_{\vec{r}_2}U(r_{12})\cdot\nabla_{\vec{p}_2}]{f}_2{d}^3\vec{p}_1{d}^3\vec{r}_2{d}^3\vec{p}_2$$
\begin{equation}
=-{{N^2{q}_1{q}_2}\over{cmV^2_0}}\int{f}_2{{\vec{r}_{12}}\over{r^2_{12}}}\cdot\nabla_{\vec{r}_1}{U}(r_{12}){d}^3\vec{p}_1{d}^3\vec{r}_2{d}^3\vec{p}_2=\nabla\cdot\vec{J}_{s3}
\end{equation}
The transformation in the right side of Eq.(67) means that $\vec{r}_1\rightarrow\vec{r}_1+(1-\lambda)\vec{r}''_{12}$, $\vec{r}_2\rightarrow\vec{r}_1-\lambda\vec{r}''_{12}$, so we have $\vec{r}_{12}=\vec{r}_1-\vec{r}_2\rightarrow\vec{r}''_{12}$ and get 
$$\vec{J}_{s3}={{q_1{q}_2{N}^2}\over{2cmV^2_0}}\int^1_0{d}\lambda\int{{\vec{r}''_{12}\cdot\vec{r}''_{12}\vec{r}''_{12}}\over{r''^2_{12}{r}''_{12}}}{{dU(r''_{12})}\over{dr''_{12}}}{f}_2[\vec{r}_1+(1-\lambda)\vec{r}''_{12},\vec{p}_1,\vec{r}_1-\lambda\vec{r}''_{12},\vec{p}_2,t]d^3\vec{r}''_{12}{d}^3\vec{p}_1{d}^3\vec{p}_2$$
\begin{equation}
={{q_1{q}_2{N}^2}\over{2cmV^2_0}}\int^1_0{d}\lambda\int{{\vec{r}''_{12}}\over{r''_{12}}}{{dU(r''_{12})}\over{dr''_{12}}}{f}_2[\vec{r}_1+(1-\lambda)\vec{r}''_{12},\vec{p}_1,\vec{r}_1-\lambda\vec{r}''_{12},\vec{p}_2,t]d^3\vec{r}''_{12}{d}^3\vec{p}_1{d}^3\vec{p}_2
\end{equation}
By using Eq.(98) and relation $\nabla_{\vec{r}_1}{U}(r_{12})=-\nabla_{\vec{r}_2}{U}(r_{12})$ again, The fifth item in the left side of Eq.(87) becomes
$$-{{N^3}\over{2V^2_0}}\int{U}_1[\nabla_{\vec{r}_1}{U}(r_{13})\cdot\nabla_{\vec{p}_1}+\nabla_{\vec{r}_2}{U}(r_{23})\cdot\nabla_{\vec{p}_2}]{f}_3{d}^3\vec{p}_1{d}^3\vec{r}_2{d}^3\vec{p}_2{d}^3\vec{r}_3{d}^3\vec{p}_3$$
\begin{equation}
={{q_1{q}_2{N}^3}\over{4cmV^2_0}}\int{f}_3{{\vec{r}_{12}}\over{r^2_{12}}}\cdot[\nabla_{\vec{r}_1}{U}(r_{13})-\nabla_{\vec{r}_2}{U}(r_{23})]{d}^3\vec{p}_1{d}^3\vec{r}_2{d}^3\vec{p}_2{d}^3\vec{r}_3{d}^3\vec{p}_3=\nabla\cdot(\vec{J}_{s3}+\vec{J}_{s4})
\end{equation}
$$\vec{J}_{s4}=-{{q_1{q}_2{N}^2}\over{8cmV^2_0}}\int^1_0{d}\lambda\int{{\vec{r}''_{13}\vec{r}''_{13}\cdot(\vec{r}_1+(1-\lambda)\vec{r}''_{13}-\vec{r}_2)}\over{r''^2_{13}(\vec{r}_1+(1-\lambda)\vec{r}''_{13}-\vec{r}_2)^2}}{{dU(r''_{13})}\over{dr''_{13}}}$$
\begin{equation}
\times{f}_3[\vec{r}_1+(1-\lambda)\vec{r}''_{13},\vec{p}_1,\vec{r}_2,\vec{p}_2,\vec{r}_1-\lambda\vec{r}''_{13},\vec{p}_3,t]d^3\vec{r}''_{13}{d}^3\vec{p}_1{d}^3\vec{p}_2{d}^3\vec{r}_3{d}^3\vec{p}_3
\end{equation}
$$\vec{J}_{s5}={{q_1{q}_2{N}^2}\over{8cmV^2_0}}\int^1_0{d}\lambda\int{{\vec{r}''_{23}\vec{r}''_{23}\cdot[\vec{r}_1-\vec{r}_2-(1-\lambda)\vec{r}''_{23}]}\over{r''^2_{23}[\vec{r}_1-\vec{r}_2-(1-\lambda)\vec{r}''_{23}]^2}}{{dU(r''_{23})}\over{dr''_{23}}}$$
\begin{equation}
\times{f}_3[\vec{r}_1,\vec{p}_1,\vec{r}_2+(1-\lambda)\vec{r}''_{23},\vec{p}_2,\vec{r}_2-\lambda\vec{r}''_{23},\vec{p}_3,t]d^3\vec{r}''_{23}{d}^3\vec{p}_1{d}^3\vec{p}_2{d}^3\vec{r}_3{d}^3\vec{p}_3
\end{equation}
Similar to Eq.(99), the first item in the right side of Eq.(87) is
$$-{{N^2}\over{2V^2_0}}\int{U}_1[\nabla_{\vec{p}_1}\cdot(\vec{G}'_{12}{f}_2)+\nabla_{\vec{p}_2}\cdot(\vec{G}'_{21}{f}_2)]{d}^3\vec{p}_1{d}^3\vec{r}_2{d}^3\vec{p}_2$$
\begin{equation}
=-{{N^2{q}_1{q}_2}\over{4cmV^2_0}}\int{f}_2{{\vec{r}_{12}\cdot{r}_{12}(\vec{G}'_{12}-\vec{G}'_{21})}\over{r^3_{12}}}{d}^3\vec{p}_1{d}^3\vec{r}_2{d}^3\vec{p}_2=\nabla\cdot\vec{J}_{s6}
\end{equation}
Similar to Eq.(98), we have
$$\vec{J}_{s6}=-{{N^2}\over{4cmV^2_0}}\int^1_0{d}\lambda\int\vec{r}''_{12}\vec{r}''_{12}\cdot[\vec{G}'_{12}(\vec{r}_1+(1-\lambda)\vec{r}''_{12},t)-\vec{G}'_{21}(\vec{r}_1-\lambda\vec{r}''_{12},t)]{{dU(r''_{12})}\over{dr''_{12}}}$$
\begin{equation}
\times{r}''_{12}{f}_2(\vec{r}_1+(1-\lambda)\vec{r}''_{12},\vec{p}_2,\vec{r}_1-\lambda\vec{r}''_{12},\vec{p}_2,t)d^3{r}''_{12}{d}^3\vec{p}_1{d}^3\vec{p}_2
\end{equation}
The second item in the right side of Eq.(87) is
$${{N^2}\over{2V^2_0}}\int{U}_1(\vec{K}_{12}\cdot\nabla_{\vec{p}_1}+\vec{K}_{21}\cdot\nabla_{\vec{p}_2}){f}_2{d}^3\vec{p}_1{d}^3\vec{r}_2{d}^3\vec{p}_2$$
\begin{equation}
={{N^2{q}_1{q}_2}\over{4cmV^2_0}}\int{f}_2{{\vec{r}_{12}\cdot{r}_{12}(\vec{K}_{12}-\vec{K}_{21})}\over{r^3_{12}}}{d}^3\vec{p}_1{d}^3\vec{r}_2{d}^3\vec{p}_2=-\nabla\cdot\vec{J}_{s7}
\end{equation}
$$\vec{J}_{s7}=-{{N^2}\over{4cmV^2_0}}\int^1_0{d}\lambda\int\vec{r}''_{12}\vec{r}''_{12}\cdot(\vec{K}_{12}-\vec{K}_{21}){{dU(r''_{12})}\over{dr''_{12}}}$$
\begin{equation}
\times{f}_2(\vec{r}_1+(1-\lambda)\vec{r}''_{12},\vec{p}_2,\vec{r}_1-\lambda\vec{r}''_{12},\vec{p}_2,t)d^3\vec{r}''{d}^3\vec{p}_1{d}^3\vec{p}_2
\end{equation}
In the formula, we should let $\vec{r}_1\rightarrow\vec{r}_1+(1-\lambda)\vec{r}''_{12}$, $\vec{r}_2\rightarrow\vec{r}_1-\lambda\vec{r}''_{12}$ in the function $\vec{K}_{12}$ and $\vec{K}_{21}$. Similarly, the third item is
$$-{{N^2}\over{2V^2_0}}\int{U}_1[(\vec{K}_{13}+\vec{F}'_{13})\cdot\nabla_{\vec{r}_1}+(\vec{K}_{23}+\vec{F}'_{23})\cdot\nabla_{\vec{p}_2}]{f}_3{d}^3\vec{p}_1{d}^3\vec{r}_2{d}^3\vec{p}_2{d}^3\vec{r}_3{d}^3\vec{p}_3$$
\begin{equation}
=-{{q_1{q}_2{N}^2}\over{4cmV^2_0}}\int{f}_2{{\vec{r}_{12}\cdot{r}_{12}(\vec{K}_{13}+\vec{F}'_{13}-\vec{K}_{23}-\vec{F}'_{23})}\over{r^3_{12}}}{d}^3\vec{p}_1{d}^3\vec{r}_2{d}^3\vec{p}_2{d}^3\vec{r}_3{d}^3\vec{p}_3=\nabla\cdot\vec{J}_{s8}
\end{equation}
$$\vec{J}_{s8}={{N^2}\over{8cmV^2_0}}\int^1_0{d}\lambda\vec{r}''_{12}\vec{r}''_{12}\cdot(\vec{K}_{13}+\vec{F}'_{13}-\vec{K}_{23}-\vec{F}'_{23}){{dU(r''_{12})}\over{dr''_{12}}}$$
\begin{equation}
\times{f}_2[\vec{r}_1+(1-\lambda)\vec{r}''_{12},\vec{p}_2,\vec{r}_1-\lambda\vec{r}''_{12},\vec{p}_2,t]d^3\vec{r}''_{12}{d}^3\vec{p}''_1{d}^3\vec{p}_2
\end{equation}
We should also let $\vec{r}_1\rightarrow\vec{r}_1+(1-\lambda)\vec{r}''_{12}$, $\vec{r}_2\rightarrow\vec{r}_1-\lambda\vec{r}''_{12}$ in $\vec{K}_{13}$, $\vec{K}_{23}$, $\vec{F}'_{13}$ and $\vec{F}'_{23}$. For the last item in the right side of Eq.(87), we have
$$-{{N^3}\over{2V^2_0}}\int{U}_1[\nabla_{\vec{p}_1}\cdot(\vec{G}'_{13}{f}_3)+\nabla_{\vec{p}_2}\cdot(\vec{G}'_{23}{f}_3)]{d}^3\vec{p}_1{d}^3\vec{r}_2{d}^3\vec{p}_2{d^3}\vec{r}_3{d}^3\vec{p}_3$$
\begin{equation}
=-{{q_1{q}_2{N}^3}\over{4cmV^2_0}}\int{f}_2{{\vec{r}_{12}\cdot{r}_{12}(\vec{G}'_{13}+\vec{G}'_{23})}\over{r^3_{12}}}{d}^3\vec{p}_1{d}^3\vec{r}_2{d}^3\vec{p}_2{d}^3\vec{r}_3{d}^3\vec{p}_3=\nabla_{\vec{r}_1}\cdot\vec{J}_{s9}
\end{equation}
$$\vec{J}_{s9}=-{{N^2}\over{4cmV^2_0}}\int^1_0{d}\lambda\int\vec{r}''_{12}\vec{r}''_{12}\cdot(\vec{G}'_{13}+\vec{G}'_{23}){{dU(r''_{12})}\over{dr''_{12}}}$$
\begin{equation}
\times{f}_2[\vec{r}_1+(1-\lambda)\vec{r}''_{12},\vec{p}_2,\vec{r}_1-\lambda\vec{r}''_{12},\vec{p}_2,t]d^3\vec{r}''_{12}{d}^3\vec{p}_1{d}^3\vec{p}_2
\end{equation}
Let $\vec{J}_s=\vec{J}_{s1}+\vec{J}_{s2}+\vec{J}_{s3}+\vec{J}_{s4}+\vec{J}_{s5}+\vec{J}_{s6}+\vec{J}_{s7}+\vec{J}_{s8}+\vec{J}_{s9}$, Eq. (87) can be written as at last
\begin{equation}
{{\partial(\rho_0{u}_s)}\over{\partial{t}}}+\nabla\cdot(\rho_0{u}_s\vec{V}+\vec{J}_s)=\rho_0\sigma_s
\end{equation}
\par
Last let's discuss interaction caused by radiation damping forces. Suppose that the charges of particles are distributed with spherical symmetry, we have 
\begin{equation}
U''(\vec{r}_1,\vec{p}_1,t)={{2q^2\dot{\vec{a}_1}\cdot\vec{p}_1}\over{3c^2m}}
\end{equation}
Similarly, producing eq.(50) by $NU''(\vec{r}_1,\vec{p}_1,t)/2V_0$ and taking integral about $d^3\vec{p}_1$, we get
$${{\partial}\over{\partial{t}}}{{N^2}\over{2V^2_0}}\int{U}''{f}_1{d}^3\vec{p}_1+{{N^2}\over{2V^2_0}}\int{U}''{{\vec{p}_1}\over{m}}\cdot\nabla_{\vec{r}_1}{f}_1{d}^3\vec{p}_1+{{N^2}\over{2V^2_0}}\int{U}''\vec{F}_{e1}\cdot\nabla_{\vec{p}_1}{f}_1{d}^3\vec{p}_1$$
$$-{{N^3}\over{2V^2_0}}\int{U}''\nabla_{\vec{r}_1}{U}(r_{12})\cdot\nabla_{\vec{p}_1}{f}_2{d}^3\vec{p}_1{d}^3\vec{r}_2{d}^3\vec{p}_2=-{{N^2}\over{2V^2_0}}\int{U}''\nabla_{\vec{p}_1}\cdot(\vec{G}'_{12}{f}_2){d}^3\vec{p}_1{d}^3\vec{r}_2{d}^3\vec{p}_2$$
\begin{equation}
-{{N^3}\over{2V^2_0}}\int{U}''(\vec{k}_{12}+\vec{F}'_{12})\cdot\nabla_{\vec{p}_1}{f}_2{d}^3\vec{p}_1{d}^3\vec{r}_2{d}^3\vec{p}_2
\end{equation}
Let energy density and energy fluid density of radiation damping force of unite mass are 
\begin{equation}
u_f={{N^2}\over{2\rho_0{V}^2_0}}\int{U}''{f}_1{d}^3\vec{p}_1
\end{equation}
\begin{equation}
\vec{J}_{f1}={1\over{2}}{{N^2}\over{V^2_0}}\int{{\vec{p}_1-m\vec{V}}\over{m}}{U}''{f}_1{d}^3\vec{p}_1
\end{equation}
We get
\begin{equation}
\rho_0{u}_f\vec{V}={1\over{2}}{{N^2}\over{V^2_0}}\int\vec{V}{U}''{f}_2{d}^3\vec{p}_1{d}^3\vec{r}_2{d}^3\vec{p}_2
\end{equation}
$$\nabla_{\vec{r}_1}\cdot(\rho_0{u}_f\vec{V}+\vec{J}_{f1})=\nabla_{\vec{r}_1}\cdot\{{1\over{2}}({N\over{V_0}})^2\int{{\vec{p}_1}\over{m}}{U}''{f}_1{d}^3\vec{p}_1\}$$
\begin{equation}
={1\over{2}}{{N^2}\over{V^2_0}}\int(U''{{\vec{p}_1}\over{m}}\cdot\nabla_{\vec{r}_1}{f}_1+f_1{{\vec{p}_1}\over{m}}\cdot\nabla_{\vec{r}_1}{U}'')d^3\vec{p}_1
\end{equation}
Therefore, the second item on the left side of Eq.(116) can be written as
\begin{equation}
{{N^2}\over{2mV^2_0}}\int{U}''\vec{p}_1\cdot\nabla_{\vec{r}_1}{f}_1{d}^3\vec{p}_1=\nabla_{\vec{r}_1}\cdot(\rho_0{u}_f\vec{V}+\vec{J}_{f1})-{1\over{2}}{{N^2}\over{2mV^2_0}}\int{f}_1\vec{p}_1\cdot\nabla_{\vec{r}_1}{U}''{d}^3\vec{p}_1
\end{equation}
Here
\begin{equation}
\nabla_{\vec{r}_1}{U}''(\vec{r}_1,\vec{p}_1,t)={{2q^2}\over{3c^2m}}\nabla_{\vec{r}_1}(\dot{\vec{a}_1}\cdot\vec{p}_1)={{2q^2}\over{3c^2m}}[\vec{p}_1\times(\nabla_{\vec{r}_1}\times\dot{\vec{a}})+(\vec{p}_1\cdot\nabla_{\vec{r}_1})\dot{\vec{a}_1}]
\end{equation}
Let
\begin{equation}
{{q^2{N}^2}\over{3c^2m^2V^2_0}}\int{f}_1\vec{p}_1\cdot[\vec{p}_1\times(\nabla_{\vec{r}_1}\times\dot{\vec{a}})+(\vec{p}_1\cdot\nabla_{\vec{r}_1})\dot{\vec{a}_1}]d^3\vec{p}_1=\rho_0\sigma_{f1}
\end{equation}
By using boundary condition, the second item in the right side of Eq.(116) can be written as
\begin{equation}
{{N^2}\over{2V^2_0}}\int{f}_1\vec{F}_{e1}\cdot\nabla_{\vec{p}_1}{U}''{d}^3\vec{p}_1={{q^2{N}^2}\over{3c^2{m}^2{V}^2_0}}\int{f}_1\vec{F}_{e1}\cdot[\vec{p}_1\times(\nabla_{\vec{p}_1}\times\dot{\vec{a}})+(\vec{p}_1\cdot\nabla_{\vec{p}_1})\dot{\vec{a}}_1]d^3\vec{p}_1=\rho_0\sigma_{f2}
\end{equation}
The forth item on the left side of Eq.(116) can also be written as
$$\nabla\cdot\vec{J}_{f2}={{q^2N^2}\over{3c^2m^2V^2_0}}\int{f}_2\nabla_{\vec{r}_1}{U}(r_{12})\cdot\nabla_{\vec{p}_1}{U}''{d}\vec{p}_1{d}^3\vec{r}_2{d}^3\vec{p}_2$$
\begin{equation}
={{q^2{N}^2}\over{3c^2m^2V^2_0}}\int{f}_2\nabla_{\vec{r}_1}{U}(r_{12})\cdot[\vec{p}_1\times(\nabla_{\vec{p}_1}\times\dot{\vec{a}})+(\vec{p}_1\cdot\nabla_{\vec{p}_1})\dot{\vec{a}}_1]d^3\vec{p}_1{d}^3\vec{r}_2{d}^3\vec{p}_2
\end{equation}
$$\vec{J}_{f2}=-{{q^2{N}^2}\over{6c^2mV^2_0}}\int^1_0{d}\lambda\int{d}^3{r}''_{12}{{\vec{r}''_{12}\vec{r}''_{12}\cdot[\vec{p}_1\times(\nabla_{\vec{p}_1}\times\dot{\vec{a}})+(\vec{p}_1\cdot\nabla_{\vec{p}_1})\dot{\vec{a}_1}]}\over{r''_{12}}}{{dU(r''_{12})}\over{dr''_{12}}}$$
\begin{equation}
\times{f}_2[\vec{r}_1+(1-\lambda)\vec{r}''_{12},\vec{p}_1,\vec{r}_1-\lambda\vec{r}''_{12},\vec{p}_2,t]d^3\vec{r}''_{12}{d}^3\vec{p}_1{d}^3\vec{p}_2
\end{equation}
Similarly, the first and second items on the right side of Eq.(116) can be written as
$$\rho_0\sigma_{f3}={{N^2}\over{2V^2_0}}\int{f}_2\vec{G}_{12}\cdot\nabla_{\vec{p}_1}{U}''{d}^3\vec{p}_1{d}^3\vec{r}_2{d}^3\vec{p}_2$$
\begin{equation}
={{N^2}\over{3c^2m^2V^2_0}}\int{f}_2\vec{G}_{12}\cdot[\vec{p}_1\times(\nabla_{\vec{p}_1}\times\dot{\vec{a}})+(\vec{p}_1\cdot\nabla_{\vec{p}_1})\dot{\vec{a}}_1]d^3\vec{p}_1{d}^3\vec{r}_2{d}^3\vec{p}_2=\sigma_{f3}
\end{equation}
$$\rho_0\sigma_{f4}={{N^3}\over{2V^2_0}}\int{f}_2(\vec{K}_{12}+\vec{F}'_{12})\cdot\nabla_{\vec{p}_1}{U}''{d}^3\vec{p}_1{d}^3\vec{r}_2{d}^3\vec{p}_2$$
\begin{equation}
={{q^2N^3}\over{3c^2mV^2_0}}\int{f}_2(\vec{K}_{12}+\vec{F}'_{12})\cdot[\vec{p}_1\times(\nabla_{\vec{p}_1}\times\dot{\vec{a}})+(\vec{p}_1\cdot\nabla_{\vec{p}_1})\dot{\vec{a}}_1]d^3\vec{p}_1{d}^3\vec{r}_2{d}^3\vec{p}_2
\end{equation}
Let $J_f=\vec{J}_{f1}+\vec{J}_{f2}$, $\sigma_f=\sigma_{f1}+\sigma_{f2}+\sigma_{f3}+\sigma_{f4}$, the energy equivalent equation of radiation damping forces is
\begin{equation}
{{\partial(\rho_0{u}_f)}\over{\partial{t}}}+\nabla\cdot(\rho_0{u}_f\vec{V}+\vec{J})=\rho_0\sigma_f
\end{equation}
At last, the energy transition equation of hydromechanics can be written as
\begin{equation}
{{\partial(\rho_0{u})}\over{\partial{t}}}+\nabla\cdot(\rho_0{u}\vec{V}+\vec{J})=-\vec{\vec{P}}:\nabla\vec{V}+\rho_0\sigma
\end{equation}
In the formula, $u=u_k+u_v+u_s+u_f$, $\vec{J}=\vec{J}_k+\vec{J}_v+\vec{J}_s+\vec{J}_f$, $\vec{\vec{P}}=\vec{\vec{P}}_k+\vec{\vec{P}}_v+\vec{\vec{P}}_s+\vec{\vec{P}}_f$, $\sigma=\sigma_k+\sigma_s+\sigma_f$. How to obtain the transport parameters from Eq.(130) remains to be researched later.
\\
\\
{\large 7. The definition of non-equivalent entropy of general systems }\\
\par
    As well known, though the thermodynamics of equilibrium states had matured since the last stage of nineteenth century, the general theory of the non-equilibrium systems has not yet been established up to now. The key to establish the general thermodynamic theory for the non-equilibrium systems is to define the correct non-equilibrium entropy. But this is still an unsolved problem now. Though some special theories for special non-equilibrium processes have defined their non-equilibrium entropies, for example, the irreversible thermodynamics based on the hypothesis of local equilibrium $^{(6)}$, the extended thermodynamics $^{(7)}$, the rational thermodynamics $^{(8)}$, as well as the famous Boltzmann's non-equilibrium entropy, none of them are with general significance.
\par
In the equilibrium thermodynamic, the first law of thermodynamic or the law of energy conservation is
\begin{equation}
dQ=dE+PdV=dE-\vec{F}\cdot{d}\vec{r}
\end{equation}
In the formula, $Q$ is total heat, $E$ is total internal energy, $P$ is pressure, $V$ is volume and $F$ is external force. For a single component system, the entropy function $S$ is defined as
\begin{equation}
TdS=dE+PdV=dE-\vec{F}\cdot{d}\vec{r}=dQ
\end{equation}
Because the concept of absolute temperature is based on equilibrium states, Eq.(132) is only suitable for equilibrium states. According to irreversible thermodynamics based on the hypothesis of local equilibrium, a system can be divided into many small cells, and each cell can be regarded as a small equilibrium system in which local equilibrium temperature $T(\vec{r},t)$ can still be defined. Let $S_m(\vec{r},t)$, $E_m(\vec{r},t)$, $Q_m(\vec{r},t)$ and $E_m(\vec{r},t)$ represent entropy density, internal energy density, heat density and external force density individually, the total entropy $S$, total internal energy $E$, total heat $Q$ and total external force $F$ of a system can be written as
\begin{equation}
S(t)=\int{S}_m(\vec{r},t){d}^3\vec{r}~~~~~~~~~~~~~~~~~~~E(t)=\int{E}_m(\vec{r},t){d}^3\vec{r}
\end{equation}
\begin{equation}
Q(t)=\int{Q}_m(\vec{r},t){d}^3\vec{r}~~~~~~~~~~~~~~~~~~~\vec{F}(t)=\int\vec{F}_m(\vec{r},t){d}^3\vec{r}
\end{equation}
The first law of thermodynamics and the definition of entropy can be written as
\begin{equation}
dQ_m=dE_m-\vec{F}_m\cdot{d}\vec{r}
\end{equation}
\begin{equation}
TdS_m=dQ_m=dE_m-\vec{F}_m\cdot{d}\vec{r}
\end{equation}
As we known that if the interactions between micro-particles are known, the ensemble probability functions can be obtained. If the ensemble probability function is known, the heat density or internal energy density is also known. In this case, the local equilibrium entropy depends on the local equilibrium temperature. In the current theory of local equilibrium thermodynamics, the function form of local equilibrium temperature $T(\vec{r},t)$ can't be determined by theory, but we can decide it through measurement point by point in principle by means of a small enough thermometer. Then we can determine the function form non-equilibrium entropy according to Eq.(136).
\par
   Because the concepts of temperature and entropy are defined based on the concept of equilibrium, when the hypothesis of local equilibrium can't hold, there are argument about whether or not the non-equilibrium temperature and non-equilibrium entropy can exit at present, thought the concepts of external energy, heat, and force are meaningful. It should be seen that the concept of entropy was put forward only for the purports to describe irreversibility of macro-system's evolution, so that the second law of thermodynamics can be expressed in the clear form of mathematics. So it is necessary for us to define the non-equilibrium entropy form this angle. The problem is how to define it, instead of whether or not it exists. For the definition of non-equilibrium entropy, two conditions should be satisfied. The first is that when a system reaches equilibrium state, the non-equilibrium entropy should identify with the equilibrium entropy defined in the equilibrium thermodynamics. The second is that the definition should satisfy the principle of entropy increment in the non-equilibrium processes. Conversely, if a function can satisfy these two conditions, it can be regarded as non-equilibrium entropy.
\par
On the other hand, entropy is not a physical quantity that can be measured directly. T the concept of entropy depends on other quantities by means of them entropy is constructed. If the concepts to construct entropy are meaningful, entropy is also meaningful. Entropy is also a extensive quantity with additivity. So we can define non-equilibrium entropy density, just as we can define energy density, heat density and forces density in equilibrium states without considering equilibrium concept.
\par
As for non-equilibrium temperature, the situation is different. Temperature is a quantity that can be measured directly. The concept of temperature depends on equilibrium closely. If equilibrium does not exist, there is no the concept of temperature, for we can not use thermometer to measure temperature in a non-equilibrium system in which the hypothesis of local equilibrium loses its meaning. Meanwhile, temperature is strength quantity. We can't define temperature density. The concept of temperature always reflects the whole nature of big enough system. So we can say that the concept of non-equilibrium temperature is meaningless while the concept of local equilibrium does not hold.
\par
In this way, we do not use the concept of non-equilibrium temperature in the general non-equilibrium systems. In order to define non-equilibrium entropy, similar to the local equilibrium entropy and by the simplest method, we introduce an unknown function $R(x,t)$ and define non-equilibrium entropy by following relation for a single component system
\begin{equation}
RdS_m=dQ_m=dE_m-\vec{F}_m\cdot{d}\vec{r}
\end{equation}
Because $R$ is not temperature, it can't be determined by the point to point's measurement in system. It is proved blow that we can decide the form of $R$ function by the method of statistical mechanics, and prove the entropy increment principle of general non-equivalent systems by the connection of statistical mechanics and thermodynamic. When systems reach equilibrium states, we have $R=T=$ constant. So the function $S_m$ defined in Eq.(137) satisfies two conditions mentioned above and can be used as the definition of non-equilibrium entropy of general systems. It can also be proved at last that the forms of non-equilibrium entropy and the function $R$ are not unique, so this is also a reason why $R$ can not be regarded as non-equilibrium temperature. Therefore, according to Eq.(137), we have
\begin{equation}
S_{m1}(\vec{r},t_1)-S_{m0}(\vec{r},t_0)=\int^{C_1}_{C_0}{{dQ_m(\vec{r},t)}\over{R(\vec{r},t)}}=\int^{C_1}_{C_0}{{1}\over{R(\vec{r},t)}}{{dQ_m(\vec{r},t)}\over{dt}}{d}t=\int^{t_1}_{t_0}{W}(\vec{r},t)dt
\end{equation}
Here $W(\vec{r},t)=\dot{Q}_m(\vec{r},t)/{R}(\vec{r},t)$, $\dot{Q}_m(\vec{r},t)=\partial{Q}_m{/}\partial{t}+\nabla{Q}_m\cdot{d}\vec{r}{/}{d}t$. It means that when the system evolves from the state $C_0$ at moment $t_0$, it reaches the state $C_1$ at moment $t_1$. By using Eq.(133), the non-equilibrium entropy's increase is
\begin{equation}
S_1(t_1)-S_0(t_0)=\int\int{{dQ_m(\vec{r},t)}\over{R(\vec{r},t)}}{d}^3\vec{r}=\int[S_{m1}(\vec{r},t_1)-S_{m0}(\vec{r},t_0)]d^3\vec{r}
\end{equation}
Now let's discuss how to decide the form of function $R$. Let $f_1(x_1,p_1,t)=f(x,p,t)$, satisfies Eq.(50). In the 6-dimension phase space, the non-equilibrium statistical entropy density is written as $S_N(\vec{r}_1\cdot\cdot\cdot\vec{r}_N,\vec{p}_1\cdot\cdot\cdot\vec{p}_N,t)$. In the 6-dimension phase space, the non-equilibrium statistical entropy density is written as  $S_1(x_1,p_1,t)=S_p(x,p,t)$. We define their relation as
\begin{equation}
S_p(\vec{r},\vec{p},t)=f^{-1}\int{f}_N(\vec{r}_1\cdot\cdot\cdot\vec{r}_N,\vec{p}_1\cdot\cdot\cdot\vec{p}_N,t)S_N(\vec{r}_1\cdot\cdot\cdot\vec{r}_N,\vec{p}_1\cdot\cdot\cdot\vec{p}_N,t)d\Omega_2\cdot\cdot\cdot\Omega_N
\end{equation}
The total entropy is
$$S(t)=\int{f}_N(\vec{r}_1\cdot\cdot\cdot\vec{r}_N,\vec{p}_1\cdot\cdot\cdot\vec{p}_N,t)S_N(\vec{r}_1\cdot\cdot\cdot\vec{r}_N,\vec{p}_1\cdot\cdot\cdot\vec{p}_N,t)d\Omega_1\cdot\cdot\cdot\Omega_N$$
\begin{equation}
=\int{f}(\vec{r},\vec{p},t)S_p(\vec{r},\vec{p},t)d^3\vec{r}{d}^3\vec{p}
\end{equation}
Comparing Eq.(144) with (133), the relation between non-equilibrium thermodynamic entropy density $S_m$ and statistical entropy density $S_p$ is
\begin{equation}
S_m(\vec{r},t)=\int{f}(\vec{r},\vec{p},t)S_p(\vec{r},\vec{p},t)d^3\vec{p}
\end{equation}
For non-equivalent states, let $dS=dS_i+dS_e$. $dS_i$ is the entropy generation in the inside of the system, $dS_e$ is the entropy fluid from the outside to the inside of the system with relations ${^(6)}$ 
\begin{equation}
{{dS_i}\over{dt}}=\int\sigma_m{d}^3\vec{r}~~~~~~~~~~~~~{{dS_e}\over{dt}}=-\int\vec{j}_m\cdot{d}\vec{\Sigma}
\end{equation}
The equilibrium equation of entropy is
\begin{equation}
{{\partial{S}_m}\over{\partial{t}}}+\nabla\cdot\vec{j}_m=\sigma_m
\end{equation}
Similar to the local equilibrium theory, for a single component system, the entropy fluid and entropy generation can also be written as
\begin{equation}
\vec{j}_m=S_m\vec{V}+{1\over{R}}\vec{J}~~~~~~~~~~~~~~\sigma_m=\vec{J}\cdot\nabla{1\over{R}}-{1\over{R}}\vec{\vec{\Pi}}:\nabla\vec{V}
\end{equation}
in the formula, $\vec{J}=\vec{J}_k+\vec{J}_v+\vec{J}_s+\vec{J}_f$ is the hot fluid discussed above, $\vec{\vec{\Pi}}$ is viscosity stress tensor. By using Eq.(50) and (141), we get
$${{dS(t)}\over{dt}}=\int({{\partial{f}}\over{\partial{t}}}{S}_p+f{{\partial{S}_p}\over{\partial{t}}})d^3\vec{r}{d}^3\vec{p}=\int(-S_p{{\vec{p}}\over{m}}\cdot\nabla_{\vec{r}}{f}+S_p{K}+f{{\partial{S}_p}\over{\partial{t}}})d^3\vec{r}{d}^3\vec{p}$$
\begin{equation}
=\int\{\int[-\nabla_{\vec{r}}\cdot(S_p{{\vec{p}}\over{m}}{f})+f{{\vec{p}}\over{m}}\cdot\nabla_{\vec{r}}{S}_p+S_p{K}+f{{\partial{S}_p}\over{\partial{t}}}]d^3\vec{p}\}d^3\vec{r}
\end{equation}
In the formula
\begin{equation}
K=-\vec{F}_{e1}\cdot\nabla_{\vec{p}_1}{f}_1-N\int(\vec{K}_{12}+\vec{F}'_{012}+\vec{F}'_{12})\cdot\nabla_{\vec{p}_1}{f}_s{d}^3\vec{r}_2{d}^3\vec{p}_s-N\int\nabla_{\vec{p}_1}\cdot(\vec{G}_{12}{f}_2){d}^3\vec{r}_2{d}^3\vec{p}_2
\end{equation}
Eq.(146) can be written as
\begin{equation}
{{\partial{S}_m}\over{\partial{t}}}=\int[-\nabla_{\vec{r}}\cdot(S_p{{\vec{p}}\over{m}}{f})+f{{\vec{p}}\over{m}}\cdot\nabla_{\vec{r}}{S}_p+S_p{K}+f{{\partial{S}_p}\over{\partial{t}}}]d^3\vec{p}
\end{equation}
Comparing it withEq.(144), we have
\begin{equation}
\vec{j}_m=\int{f}{{\vec{p}}\over{m}}S{d}^3\vec{p}~~~~~~~~~~~~~\sigma_m=\int(f{{\vec{p}}\over{m}}\cdot\nabla_{\vec{r}}{S}_p+KS_p+f{{\partial{S_p}}\over{\partial{t}}})d^3\vec{p}
\end{equation}
By using Eq.(145), we get£º
\begin{equation}
S_m\vec{V}+{1\over{R}}\vec{J}=\int{f}{{\vec{p}}\over{m}}S_p{d}^3\vec{p}
\end{equation}
\begin{equation}
\vec{J}\cdot\nabla{1\over{R}}-{1\over{R}}\vec{\vec{\Pi}}:\nabla\vec{V}=\int(f{{\vec{p}}\over{m}}\cdot\nabla_{\vec{r}}{S}_p+KS_p+f{{\partial{S}_p}\over{\partial{t}}})d^3\vec{p}
\end{equation}
Put Eq.(142) and $\vec{J}=\vec{J}_k+\vec{J}_v+\vec{J}_s+\vec{J}_f$ into Eq.(150), we can write
\begin{equation}
\vec{J}(\vec{r},t)=\int\vec{J}_q(\vec{r},\vec{p},t){d}^3\vec{p}
\end{equation}
Then replacing the integral about $d^3\vec{p}$, we get the form of the non-equilibrium statistical entropy
\begin{equation}
S_p(\vec{r},\vec{p},t)={{G(\vec{r},\vec{p},t)}\over{R(\vec{r},t)}}
\end{equation}
\begin{equation}
G(\vec{r},\vec{p},t)={1\over{f}}{{(\vec{p}-m\vec{V})}\over{(\vec{p}-m\vec{V})^2}}\cdot\vec{J}_q(\vec{r},\vec{p},t)
\end{equation}
Putting Eq.(153) into Eq.(151), we get
\begin{equation}
{{\partial{R}}\over{\partial{t}}}+\vec{A}\cdot\nabla{R}=BR
\end{equation}
In the formula
\begin{equation}
\vec{A}={{\int{f}G\vec{p}{d}^3\vec{p}}\over{m\int{f}Gd^3\vec{p}}}~~~~~~~~~~~B={{\vec{\vec{\Pi}}:\nabla\vec{V}+\int(f{{\vec{p}}\over{m}}\cdot\nabla_{\vec{r}}{G}+KG+f{{\partial{G}}\over{\partial{t}}})d^3\vec{p}}\over{\int{f}Gd^3\vec{p}}}
\end{equation}
On the other hand, the viscosity stress tensor in Eq.(145) can be written as $\vec{\vec{\Pi}}=P\delta_{ij}+\varepsilon_{ij}$, here $P$ is pressure and it is known or can be calculated in principle if $f$ is known. $\varepsilon_{ij}$ is the viscosity tensor and can also be calculated if $f$ is known. So $\vec{\vec{\Pi}}$ can be regarded as known quantity. Therefore, as long as the probability function $f$ is known by means of statistical physics, we can achieve the form of $R$ function from Eq.(155) in principle. Thus the forms of non-equilibrium statistical entropy in the 6-dimention phase space can be determined by Eq.(153) and the form of non-equilibrium thermodynamic entropy can be determined by Eq.(142). The form of non-equilibrium statistical entropy in the $6N$ -dimention phase space can also be determined by Eq.(140) in principle.
\par
The equilibrium states are discussed now. If we define the states with $R$ =constant as the equilibrium states, according to Eq.(155), the equilibrium condition is $B=0$ . From Eq.(156), by considering the fact that we have $\vec{V}$ =constant when a system reaches equilibrium states, the equilibrium condition becomes
\begin{equation}
\int(f{{\vec{p}}\over{m}}\cdot\nabla_{\vec{r}}{G}+KG+f{{\partial{G}}\over{\partial{t}}})d^3\vec{p}=0
\end{equation}
This means $\sigma_m=0$ comparing with Eq.(149) when $R$ =constant. 
\par
On the other hand, if $R$ has nothing to do with space coordinate $\vec{r}$ but relative to time $t$, we have $R=R(t)=T(t)$, $T$ is absolute temperature. Suppose $Q_m$ also has nothing to do with space coordinate $\vec{r}$ but relative to time $t$, i.e, $Q_m=Q_m(t)$, because
\begin{equation}
{{dQ_m}\over{dt}}={{\partial{Q}_m}\over{\partial{t}}}+\nabla{Q}_m\cdot{{d\vec{r}}\over{dt}}={{\partial{Q}_m}\over{\partial{t}}}
\end{equation}
Put it into Eq.(139), we get
\begin{equation}
S(t)-S_0=\int{1\over{T}}\int{{dQ_m}\over{dt}}d^3\vec{r}dt=\int{1\over{T}}{{\partial}\over{\partial{t}}}\int{Q}_m{d}^3\vec{r}dt=\int{1\over{T}}{{\partial{Q}}\over{\partial{t}}}dt=\int{{dQ}\over{T}}
\end{equation}
This is just equilibrium entropy in equilibrium thermodynamics. Therefore, the temperature can be the functions of time in the processes of equilibrium states, as long as the systems are uniform. The processes are just the so-called quasi-stationary processes, but this point has not been noted clearly in the current thermodynamics. It is useful for us to calculate entropy function by using this nature.
\par
Now let we prove the principle of entropy increment in the adiabatic processes. That is to prove the following relation for the non-equilibrium isolated systems
\begin{equation}
{{dS}\over{dt}}=\int{1\over{R}}{{dQ_m}\over{dt}}d^3\vec{r}>{0}
\end{equation}
In principle, if the distribution function is known, we can prove it directly by the method of statistical mechanics, but this is impossible at present because of difficulty in mathematics. So we prove it by the method of connecting thermodynamics and statistical mechanics.
\par
Firstly, because $Q_m(t)$ and $R(t)$ are the continuous functions of time, so $S(t)$ is also a continuous function of time. Next, suppose an isolated system is in an equilibrium state at time $t_0$. At time $t_0+dt$ a disturbing force is acted on the system so that the system becomes a non-equilibrium states. Then the disturbing force is removed immediately and the system would evolutes in the isolated state and reaches another equilibrium state at time $t_n$. It is impossible to have another equilibrium state between these two equilibrium states. If there exists the third equilibrium state, it means that the system would become non-equilibrium from equilibrium without disturbing force form outside. In this case, the second law of thermodynamics would be violated. Thus, what we should prove is that the entropy would never decrease during the non-equilibrium process between two equilibrium states. By using the method of reduction to absurdity, we will prove below that $\Delta{S}$ should be a monotonously increasing function during the whole time $t_0\rightarrow{t}_n$.
\par
According to the theory of equilibrium thermodynamics, because the system is in the equilibrium states at initial and final states, we have $\Delta{S}=S(t_n)-S(t_0)>0$. If $\Delta{S}$ does not increase monotonously, when $\Delta{S}$ changes from $dS>0$ to $dS<0$ or from $dS<0$ to $dS>0$, it would appear a state with $dS=0$ or $dS/{d}t=0$ at a certain moment $t_i$ with $t_0<t_i{<}t_n$. According to equilibrium theory, it means that the system must be in the equilibrium states at moment $t_i$. However, this is impossible as mention above. On the other hand, because $\vec{F}_m=0$ for an isolated system, we have $\dot{Q}_m=\dot{E}_m$ and get according to Eq.(138)
\begin{equation}
{{dS}\over{dt}}=\int{{\dot{Q}_m}\over{R}}{d}^3\vec{r}=\int{{\dot{E}_m}\over{R}}{d}^3\vec{r}=0
\end{equation}
There are two ways to make Eq.(161) possible. The first is $\dot{E}_m=0$ or $E_m$ =constant. However, this is impossible for $f\neq$ constant, so $E_m\neq$ constant in the non-equilibrium processes. The second way is that $\dot{E}_m{/}R$ is not a single valued function so that for any boundary condition $\vec{r}=\vec{r}_1$ and $\vec{r}=\vec{r}_2$, we always have 
\begin{equation}
\int^{\vec{r}_2}_{\vec{r}_1}{{\dot{E}_m}\over{R}}{d}^3\vec{r}=W(\vec{r}_2,t)-W(\vec{r}_1,t)=0
\end{equation}
However, this condition can't be satisfied generally, for the probability distribution function is a single valued function in general. So Eq.(161)and (162) can not hold in the non-equilibrium processes in general. Therefore, $dS/dt\neq{0}$ in non-equilibrium processes and non-equilibrium entropy $S$ must be a monotone function. On other hand, because $\Delta{S}=S(t_n)-S(t_0)>0$, so during the process from time $t_0$ to $t_n$, $S$ must be a monotonously increasing function. Thus, the principle of non-equilibrium entropy increment is proved for the non-equilibrium processes.
\par
At last we discuss the uniqueness problem of the definition of non-equilibrium entropy. Because probability distribution function $f$ is unique, the form of function $R$ is unique, so the form of non-equilibrium entropy shown in (137) is also unique. On the other hand, if there exists other form's non-equilibrium entropies $S'_m$, we can always define them as
\begin{equation}
R'dS'_m=dQ_m+dY_m
\end{equation}
If $Y_m$ is an unknown function, the definition is meaningless for there are three unknown functions and we can't decide them by the method mentioned above. If $Y_m$ is the function of known functions, for example energy density, heat density and force density and so on, we can also determinate the forms of $S'_m$ and $R'$ by the same method. If $dY_m=0$ and $R'=T$ in equilibrium states, the form of $S'_m$ would be the same as equilibrium entropy in equilibrium state. If $Y_m$ and $R'$ are also the single valued functions and $\dot{Y}_m{/}R'\neq{0}$ in non-equilibrium states, we can also prove the principle of non-equilibrium entropy increment. If all conditions are satisfied, we can also consider $S'_m$ as the non-equilibrium entropy density. On the other hand, from Eq.(137) and (163) we have
\begin{equation}
RdS=R'd(S'_m-U_m{/}R')
\end{equation}
\begin{equation}
S_m(\vec{r},t)=\int{{R'}\over{R}}{{d(S'_m-Y_m{/}R')}\over{dt}}{d}t+S_{m0}=\int{{R'}\over{R}}(\dot{S}'_m-{{\dot{Y}_m}\over{R'}}+{{Y_m\dot{R}'}\over{r'^2}})dt+S_{m0}
\end{equation}
In this way, both $S_m$ and $S'_m$ can be regarded as non-equilibrium entropy densities connecting by relation above. So the definition of non-equilibrium entropy is not unique. This is just the reason why we can define different non-equilibrium entropies in the current non-equilibrium thermodynamics. Because $R\neq{R}'$, so $R$ can't be regarded as non-equilibrium temperature, for non-equilibrium temperature as a measurable physical quantity should be unique if it exists. Because non-equilibrium entropy is directly immeasurable, so it can be non-unique. In this way, non-equilibrium entropy is not a state function and its increment can also be non-unique, thought equilibrium entropy is a state function and its increment is unique.
\par
Now we have completed the reform of classical statistical mechanics. By considering retarded electromagnetical interaction, we can also introduce asymmetry of time reversal into quantum theory. This problem will be discussed later.
\\
\\
\\
\\
Reference
\\
1. Miao Dongsheng, Liu Huajie, Great Ease On Chaos, People's University Publishing House, 262£¨1993£©.
\\
2. Wang Zhuxi, Introduction to Statistical Physics, People's Education Publishing House, 34£¬152£¨1965£©.
\\
3. S.Chandrasekhar, Rev. Mod.Phys., 15£¬84 (1943).
\\
4. Cao Changqi, Electrodynamics, People's Education Publishing House, 240£¨1979£©.
\\
5. Luo Liaofu, Theory of Non-equilibrium Statistics, Neimenggu University Publishing House, 355, 358£¨1990£©.
\\
6.De Groot and Mazur, Nonequilibrium Thermodynamics£¨1962£©. P.Glansdorff and I.Prigogine, Thermodynamics Theory of Structure, Stability and Fluctuations, £¨1971£©.
\\
7.S.Simons, J.phys. A: Math., Nucl. Gen., 6,1934(1973). G.Lebon, et al., J.phys. A, 13(1980), 275.D.Jou, J.Casas-Vazquez, G.Lebon, Rep. Prog. Phys. 51 1105(1988).
\\
8.C.Trusdell, Rational Thermodynamics,£¨1969£©. B.p.Coleman, J.Chem. Phys., 47, 597(1967).
  W.Noll,Arch. Rational Mech. Anal., 17, 85(1973).
\end{document}